\documentclass[showpacs,preprintnumbers,amsmath,amssymb]{revtex4}

\usepackage[latin1]{inputenc}
\usepackage{makeidx}
\usepackage{layout}        
\usepackage{color}
\usepackage{graphicx}
\usepackage{dcolumn}
\usepackage{bm}

\def\pp{ \|}

\newcommand{\bra}[1]{\langle #1|}
\newcommand{\ket}[1]{|#1\rangle}
\newcommand{\braket}[2]{\langle #1|#2\rangle}

\begin{document}
\preprint{preprint}

\title{Magnetism of iron: from the bulk to the monoatomic wire}

\author{     Gabriel Aut\`es$^*$, Cyrille Barreteau$^*$, Daniel Spanjaard$^{\dag}$ and Marie-Catherine Desjonqu\`eres$^*$ }
\affiliation{$^*$CEA Saclay, DSM/DRECAM/SPCSI, B\^atiment 462,
F-91191 Gif sur Yvette, France }

\affiliation{$^{\dag}$Laboratoire de Physique des Solides,
             Universit\'e Paris Sud, B\^atiment 510, F-91405 Orsay, France}

\date{\today}
\begin{abstract}
The magnetic properties of iron (spin and orbital magnetic moments, magnetocrystalline anisotropy energy)
in various geometries and dimensionalities are investigated by using a parametrized tight-binding
model in an $s$, $p$ and $d$ atomic orbital basis set including spin polarization and the effect of spin-orbit
coupling. The validity of this model is well established by comparing the results with those
obtained by using an ab-initio code. This model is applied to the study of iron in  bulk bcc and fcc phases,
$(110)$ and $(001)$ surfaces and  to the monatomic wire, at several interatomic distances.  New results
are derived. The variation of the component of the orbital magnetic moment on the spin quantization
axis has been studied as a function of depth, revealing a significant enhancement in the first two layers,
especially for the $(001)$ surface.
It is found that the magnetic anisotropy energy is drastically increased in the wire and can reach several
meV. This is also true for the orbital moment, which in addition is highly anisotropic. Furthermore it is
shown that when the spin quantization axis is neither parallel nor perpendicular to the wire the average
orbital moment is not aligned with the spin quantization axis. At equilibrium distance the  easy magnetization axis is
along the wire but switches to the perpendicular direction under compression.
The success of this model opens up the possibility of obtaining accurate results on other elements and
systems with much more complex geometries.
\end{abstract}


\maketitle

\section{Introduction}

The magnetic properties of nanoparticles, thin films and wires have
attracted recently a lot of attention due to their potential
technological applications. It is thus very important to
investigate the influence of dimensionality on these properties.
In many experimental systems, some atoms are in a bulk-like
environment while some others have a very low coordinence in a
strongly asymmetric environment. This is the case of clusters, at
the edge of adsorbed islands, supported wires, along step edges or
in the nanoconstriction region of a break junction etc.. As a
consequence the magnetic properties of such systems need to be
treated at an atomic level, by studying their electronic structure
in the framework of quantum mechanics.

In principle these properties can be determined from ab-initio
calculations. However the computer time and storage increase
drastically with the number of inequivalent atoms. Thus there is a
need for simplified methods still based on quantum mechanics which
capture the essential physics. In this context tight-binding (TB)
methods are ideally suited for these calculations. Indeed TB
methods can be extended to magnetic systems by introducing Hubbard
two-body terms treated in the Hartree-Fock approximation
\cite{Barreteau00,Barreteau04}, and can easily include spin-orbit
effects\cite{Friedel64}. Moreover they allow the calculation of
local physical quantities (spin or orbital moment etc..) in a
straightforward manner which gives a physically transparent
understanding of the phenomena.

The anisotropy of the magneto-crystalline energy (MAE), although
small in transition metals, is of fundamental importance since it
determines the easy magnetization axis. The MAE results from the
coupling of the spin and orbital moments. In the bulk it is
well known that the orbital moment is nearly quenched due to the
high symmetry of the potential. When dimensionality or symmetry is
reduced the orbital moment is less and less quenched and the MAE
increases rapidly. The MAE is routinely measured by magnetic
hysteresis or torque
measurements\cite{Stearns86,Gradmann93,Bruno89th}, for instance.
More recently the development of X-ray magnetic circular dichroism
techniques has allowed the experimental determination of the
orbital moment\cite{XMCD_book}. On the theoretical side the
calculation of the MAE in bulk pure ferromagnetic transition
metals is a challenge because of its minuteness (typically some
$\mu$eV). The first attempts to determine the MAE on surfaces have
been performed on free standing mono-layers either using a
perturbative treatment of the spin-orbit coupling in a
tight-binding model\cite{Bruno89th,Bruno89}, or with
self-consistent ab-initio technique\cite{Gay86,Wang93}. The calculations performed by
Bruno \cite{Bruno89th,Bruno89} on mono-layers have been
extended to study slabs with several atomic
layers\cite{Cinal94,Lessard97} in a pure $d$-band model. More
recently the case of layered ordered alloys of the CuAu or CsCl types containing at least
one ferromagnetic element\cite{Burkert05}, deposited ferromagnetic over-layers on
non-magnetic substrates\cite{Qian97}, and multilayers have been
investigated\cite{Kyuno96,Broddefalk02}. Simultaneously the orbital magnetism driven by
the spin-orbit interaction has been computed using the
tight-binding model (perturbatively\cite{Bruno89th,Bruno89} or non
perturbatively\cite{Guirado03,Dorantes97}) as well as ab-initio
codes\cite{Popescu01,Eriksson90}.

The aim of this paper is to develop a TB model allowing the
determination of the magnetic properties of transition metals in various
geometrical configurations, going from highly coordinated and symmetric
environments to low coordinated and anisotropic geometries. In a first step we have found
useful to check the validity of the model by a detailed comparison with the results 
provided by ab-initio methods on simple systems. In addition, the use of both methods 
yields a better physical understanding of these properties, and, moreover, the
simplicity of our model allows a more detailed analysis of each system. 
We use a non-orthogonal basis set of $s$, $p$ and $d$ valence orbitals. 
The parametrization of the non-magnetic and non-relativistic
Hamiltonian was derived by Mehl and Papaconstantopoulos \cite{Mehl96}. The possibility of
spin polarization is then introduced using a Stoner like model in which the splitting 
between the energy levels of up and down spin orbitals is governed by the Stoner
parameter $I_{dd'}$ and is proportional to the magnetic
moment carried by $d$ electrons. Indeed it is well known that $s$ and $p$ electrons
are very weakly polarized. The relativistic effects, limited to spin-orbit coupling
between $d$ electrons, are taken into account by adding the intra atomic matrix 
elements of this coupling determined by a  parameter $\xi$.
The two parameters $I_{dd'}$ and $\xi$ are fixed by comparison with ab-initio calculations. 
For the Stoner parameter the variation of the magnetization as a function of interatomic
distance in the bulk phase can be used.  The determination of the spin-orbit coupling 
parameter relies on the study of the degeneracy removal of energy bands at high symmetry points or directions of the Brillouin zone. 

We have applied our model to the study of iron in various atomic arrangements going from the bulk,
in the experimentally observed phases  bcc and fcc, to simple surfaces and finally to the
monoatomic wire, at several interatomic distances. 
It is found that with a unique value of the Stoner parameter we are able to reproduce the
variation of the spin magnetic moment  in a wide range of lattice spacings in the bcc bulk phase.
It is well known that the magnetic properties of bulk fcc iron are rather complex. In spite of 
this complexity the calculations carried out in this work are in excellent agreement with 
ab-initio predictions. The spin-orbit coupling parameter is then determined, as explained above,
and a single value of this parameter is able to reproduce the ab-initio band structure.
Furthermore the contribution of the orbital moment to the magnetization in bcc Fe is very
close to the experimental value. The case of surfaces represents a more stringent check 
since the dimensionality is reduced and all atoms are not geometrically equivalent. 
It is in particular interesting to follow the variation of the magnetic properties when
going from the outermost to inner (bulk-like) layers. The spin-polarized surface projected
band structure as well as the variation of the spin magnetic moment as a function of depth
derived from ab-initio calculations are perfectly reproduced by our model. 
In particular the $(001)$ surface atoms have a saturated moment contrary to those of the $(110)$ 
surface. Introducing the spin-orbit coupling in the TB Hamiltonian allows the calculation of 
the MAE and of the orbital magnetic moment at each inequivalent site. Due to the efficiency
of our model it is possible to check rigorously the convergence of these quantities 
when increasing the number of $k$ points. 
The orbital moment is strongly increased on surface atoms (about twice the bulk value for
the $(001)$ surface), and recovers its bulk value on the third and innermost layers.
The wire is the most anisotropic atomic arrangement with the lowest coordinence.
Even though the TB parameters are fitted on bulk ab-initio data only, the non-magnetic
band structure calculated with these parameters is in good agreement with ab-initio 
calculations, proving their good transferability. The two additional parameters $I_{dd'}$ and
$\xi$ are also perfectly transferable. Indeed, without changing the Stoner parameter, the
variation of the spin magnetic moment with the interatomic distance is satisfactorily
reproduced  (in particular the saturated solution appears abruptly at the same
interatomic spacing) and the splitting of bands due to spin-orbit coupling
is exactly the same, compared to ab-initio results.
The calculation of the MAE reveals that at theoretical equilibrium the
easy axis is parallel to the wire, but at smaller interatomic distances corresponding
to unsaturated magnetic solutions the easy axis is perpendicular to the wire.
We have finally checked the validity of Bruno formula \cite{Bruno89} relating
the MAE to the anisotropy of the orbital moment, and found that this relation
is almost strictly obeyed around the equilibrium distance. Indeed at this
interatomic distance the up spin bands are filled and the exchange splitting
is large compared to the $d$ bandwidth, which was not the case for the $(001)$
surface that, although saturated, has much a wider $d$ bandwidth. 

The paper is organized as follows. In section 2 the formalism of our model
is presented in details, in particular the derivation of the $x$, $y$ and $z$ components
of the orbital and spin moment formula in a non-orthogonal basis set. 
The spin-orbit coupling being small we have recalled the perturbation treatment
of the MAE and of the orbital moment, from which the analytical expression of the anisotropy
laws are directly obtained, and are used to analyze our numerical results.
In section 3 the Stoner parameter is determined and used to study in detail the
magnetic properties of bcc and fcc iron. Section 4 is devoted to the
study of $(001)$ and $(110)$ surfaces. Finally in section 5 we present an
exhaustive study of the monoatomic wire. Conclusions are drawn in section 6.

\section{Formalism}

\subsection{Spin polarized tight-binding model}

We choose as a basis set the real {\it s, p} and {\it d} valence
atomic orbitals centered on each site $i$. They are denoted by
$\lambda$ and $\mu$ indices ($\lambda, \mu=1,9$) and numbered as
follows: $s, p_x, p_y, p_z, d_{xy}, d_{yz}, d_{zx}, d_{x^2-y^2},
d_{3z^2-r^2}$, the $x, y, z$ coordinates being taken along the
crystal axes. The tight-binding (TB) hamiltonian for the
non-magnetic (NM) state is then completely determined by its
intra-atomic matrix elements (i.e., the {\it s, p} and {\it d}
atomic levels) $\varepsilon_{\lambda}$ and its interatomic matrix
elements (i.e., the hopping integrals) $\beta_{ij}^{\lambda\mu}$
which have been tabulated as a function of 10 Slater-Koster (SK)
parameters ($ss\sigma, sp\sigma, sd\sigma, pp\sigma, pp\pi,
pd\sigma, pd\pi, dd\sigma, dd\pi, dd\delta$) and of the direction
cosines of the bonding direction {\bf R$_{ij}$}\cite{Slater54}.
Following the (MP) scheme developed by Mehl and
Papaconstantopoulos \cite{Mehl96}, the atomic levels depend on the
atomic environment (number of neighbors and interatomic distances)
while the SK parameters are function of $R_{ij}$ only. Finally the
Schroedinger equation in the atomic orbital basis involves also
overlap integrals $\mathcal{S}_{ij}^{\lambda\mu}$ depending on 
the bonding direction $\bm{R}_{ij}$ when the non-orthogonality of the
basis set is taken into account. All these quantities
($\varepsilon_{\lambda}$, $\beta_{ij}^{\lambda\mu}$,
$\mathcal{S}_{ij}^{\lambda\mu}$) are written as analytic functions
depending on a number of parameters which are determined by a
least mean square fit of the results of ab-initio electronic
structure (band structure and total energy) calculations either in
the Local Density (LDA) or in the Generalized Gradient (GGA)
approximations. These parametrizations will be denoted as TBLDA
and TBGGA in the following. The analytical form of the functions
can be found in Ref.\cite{Mehl96} and the numerical values
of the parameters for Fe can be found in Ref.\cite{PapaWeb}.

In order to account for spin polarization, we use a simplified
Hartree-Fock (HF) scheme \cite{Barreteau00} to define atomic levels
depending on spin ($\varepsilon_{\lambda\sigma}$). These diagonal
elements of the hamiltonian can be written in the basis of
spin-orbitals $\ket{i\lambda\sigma}$ in which the spin quantization
axis is parallel to the magnetization ($\sigma=+1(-1)$ for
up(down) spin). When all atoms in the system are geometrically
equivalent we get:

\begin{eqnarray}
\varepsilon_{s,\sigma}&=&\varepsilon_{0s}+U_{ss}\frac{N_s}{2}+(U_{sp}-\frac{J_{sp}}{2})
N_p+(U_{sd}-\frac{J_{sd}}{2})N_d \nonumber \\
&-&\frac{\sigma}{2}(U_{ss}M_s+J_{sp}M_p+J_{sd}M_d) \nonumber \\
\varepsilon_{p_{\alpha}\sigma}&=&\varepsilon_{0p}+U_{sp}N_s+(U_{pp'}-\frac{J_{pp'}}{2})
N_p-\frac{1}{2}(U_{pp'}-3J_{pp'})n_{p_{\alpha}}+U_{pd}N_d  \nonumber \\
&-&\frac{\sigma}{2}(J_{sp}M_s+J_{pp'}M_p+J_{pd}M_d
+ (U_{pp'}+J_{pp'})m_{p_{\alpha}}) \nonumber \\
\varepsilon_{d_{\alpha}\sigma}&=&\varepsilon_{0d}+(U_{sd}-\frac{J_{sd}}{2})N_s+
(U_{pd}-\frac{J_{pd}}{2})N_p+(U_{dd'}-\frac{J_{dd'}}{2})N_d
-\frac{1}{2}(U_{dd'}-3J_{dd'})n_{d_{\alpha}} \nonumber \\
&-&\frac{\sigma}{2}(J_{sd}M_s+J_{pd}M_p+J_{dd'}M_d +
(U_{dd'}+J_{dd'})m_{d_{\alpha}}) \label{eq:HFlevels}
\end{eqnarray}

\noindent where $N_{s(p,d)}$ and $M_{s(p,d)}$ are the total number
of electrons and the total moment on each atom, respectively, in
{\it s, p} and {\it d} orbitals while $n_{p_{\alpha}(d_{\alpha})}$
and $m_{p_{\alpha}(d_{\alpha})}$ denote the total occupation
number (i.e., for both spins) and moment in orbital
$p_{\alpha}(d_{\alpha})$. Finally  the $U$ and $J$ parameters are
Coulomb and exchange integrals which involve two different
orbitals, save for $U_{ss}$, and are assumed to depend only on the
orbital quantum numbers of these orbitals.

We further assume that the asphericity of both the charge distribution
and magnetic polarization can be neglected, i.e.,
$n_{p_{\alpha}}=N_p/3$, $n_{d_{\alpha}}=N_d/5$,
$m_{p_{\alpha}}=M_p/3$, $m_{d_{\alpha}}=M_d/5$. In these
conditions, when the system is non magnetic all the non vanishing
terms in Eqs.\ref{eq:HFlevels} are accounted for implicitly by the
expression of $\varepsilon_{\lambda}$ in the MP scheme \cite{Mehl96}. In addition the
equations giving the atomic levels can be further simplified by
noting that the spin polarization of $s$ and $p$ electrons is very
small. As a consequence Eq.\ref{eq:HFlevels} can be approximated by:

\begin{eqnarray}
\varepsilon_{s,\sigma}=\varepsilon_{s}-\frac{\sigma}{2}J_{sd}M_d
\nonumber \\
\varepsilon_{p,\sigma}=\varepsilon_{p}-\frac{\sigma}{2}J_{pd}M_d
\nonumber \\
\varepsilon_{d,\sigma}=\varepsilon_{d}-\frac{\sigma}{2}I_{dd'}M_d
\label{eq:levels}
\end{eqnarray}

\noindent where $\varepsilon_{s(p,d)}$ are the NM levels
\cite{Mehl96} and $I_{dd'}=(U_{dd'}+6J_{dd'})/5$ can be identified
with the Stoner parameter. The numerical value of this parameter
will be determined in section 3.1 in order to reproduce as closely
as possible the variation of the magnetic moment as a function of
the interatomic distance that can be obtained from an ab-initio
calculation. Finally, $J_{sd}$ and $J_{pd}$ are one order of
magnitude smaller than $I_{dd'}$ \cite{Barreteau00} and we have
taken $J_{sd}=J_{pd}=I_{dd'}/10$. This completely defines our TB
spin polarized hamiltonian $H_{TBHF}$ in the absence of spin-orbit
coupling for a system of equivalent atoms, and when the overlaps
are neglected.

In the general case where overlaps are taken into account and all atoms
in the systems are not geometrically equivalent, the Hamiltonian becomes:

\begin{equation}
H_{ij}^{\lambda \mu \sigma} =H_{ij}^{0,\lambda \mu} +
\frac{U}{2}(\delta N_{i}+\delta N_{j})\mathcal{S}_{ij}^{\lambda \mu}
-\frac{\sigma}{4}(\Delta_{\lambda}^i+\Delta_{\mu}^j)\mathcal{S}_{ij}^{\lambda \mu}
\label{eq:Htot}
\end{equation}

\noindent
$H_{ij}^{0,\lambda \mu}$ are the matrix elements provided by the MP parametrization
of the Hamiltonian. The second term in which $\delta N_{i}$ is the net total charge on atom $i$,
prevents large charge transfers when inequivalent atoms are present, and will be discussed in section 4.
 Finally in the last term which accounts for spin polarization 
 $\Delta_{\lambda}^i=J_{sd}M_d^i,J_{pd}M_d^i$ and $I_{dd'}M_d^i$ for $s,p,d$ orbitals, respectively.

\subsection{The spin-orbit coupling}

The spin-orbit interaction for a single atom is given by:

\begin{equation}
H_{\text{so}}=\frac{\hbar}{4m^2c^2}(\mbox{\boldmath$\nabla$} V\wedge{\bf
p}).\mbox{\boldmath$ \sigma$}
\end{equation}

\noindent where $V$ is the atomic potential, ${\bf p}$ is the
momentum operator and \mbox{\boldmath$ \sigma$} are the 
Pauli matrices. Taking into account the spherical symmetry of the
potential, $H_{\text{so}}$ can be rewritten as:

\begin{equation}
H_{\text{so}}=\xi(r){\bf L}.{\bf S}
\end{equation}

\noindent with:

\begin{equation}
\xi(r)=\frac{1}{2m^2c^2} \frac{1}{r}\frac{dV}{dr}.
\end{equation}
${\bf L}={\bf r}\wedge{\bf p}$ and ${\bf S}=\hbar\bm{\sigma}/2$ are, respectively, the angular orbital and spin
momentum operators. The matrix elements of $H_{\text{so}}$ in the basis
of atomic spin-orbitals $\ket{\lambda\sigma}$ are:

\begin{equation}
\bra{\lambda\sigma}H_{\text{so}}\ket{\mu\sigma'}=\xi_{\lambda\mu}\bra{\bar{\lambda}\sigma}
{\bf L}.{\bf S}\ket{\bar{\mu}\sigma'}
\end{equation}

\noindent with:

\begin{equation}
\xi_{\lambda\mu}=\int_0^\infty\mathcal{R}_{\lambda}(r)\mathcal{R}_{\mu}(r)\xi(r)r^2dr
\end{equation}

\noindent where $\mathcal{R}_{\lambda}(r)$ is the radial part of
the atomic orbital $\lambda$ and $\bar{\lambda}$ denotes its
angular part. Since $\xi(r)$ is well localized around $\bf{r}=0$,
$\xi_{\lambda\mu}$ has a non negligible value only when
$\mathcal{R}_{\lambda}(r)$ and $\mathcal{R}_{\mu}(r)$ are also
well localized, i.e., for transition metals, when both $\lambda$
and $\mu$ are {\it d} orbitals in which case
$\mathcal{R}_{\lambda}(r)=\mathcal{R}_{\mu}(r)=\mathcal{R}_d(r)$
and $\xi_{\lambda\mu}=\xi$ ($\xi>0$).

In the tight-binding approximation the crystal potential is
written as $\mathcal{V}({\bf r})=\sum_iV(|{\bf r}-{\bf R}_i|)$ and
$H_{\text{so}}$ becomes:

\begin{displaymath}
H_{\text{so}}=\sum_i\xi(|{\bf r}-{\bf R}_i|){\bf L}_i.{\bf S}
\end{displaymath}

\noindent where ${\bf L}_i$ is the angular orbital momentum
operator with respect to the center $i$. For transition metals and
due to the localized character of $\xi(|{\bf r}-{\bf R}_i|)$ we
can neglect all matrix elements of $H_{so}$ save for the
intra-atomic ones between $d$ orbitals. These matrix elements are
the same at each site and given in Appendix A in the spin
framework $x'', y'', z''$. In this framework $z''$ is the spin
quantization axis defined by its polar  and  azimuthal angles
$\theta, \varphi$ relative to the crystal axes. The $x''$ and
$y''$ axes have been chosen in the following way: the $x, y$ axes
of the crystal are first rotated by the angle $\varphi$ around
$z$, this gives a new framework $x', y', z'$ which is then rotated
by an angle $\theta$ around $y'$. The orbital and spin moments
are usually expressed in units of $\hbar$ so that $\xi$ is a
parameter which has the dimension of an energy. Its numerical
value will be deduced from ab-initio calculations in the
following.

\subsection{Determination of the components of the spin and orbital
moments in the spin framework. }

\subsubsection{Spin moment. }

Let us first compute the average value of the three components of
the total spin $<S_{x''}>, <S_{y''}>, <S_{z''}>$ in the spin
framework. If we choose as a basis set of spin-orbitals the direct
product of the orbitals $\ket{i\lambda}$ with the eigenvectors of the
operator $S_{z''}$ denoted as $\uparrow$ and $\downarrow$, the
electron eigenfunctions $\ket{\psi_n}$ in the crystal can be written:

\begin{displaymath}
\ket{\psi_n}=\sum_{i\lambda}c_{i\lambda\uparrow}^n \ket{i\lambda\uparrow}+
c_{i\lambda\downarrow}^n\ket{i\lambda\downarrow}=
\sum_{i\lambda\sigma}c^n_{i\lambda\sigma}\ket{i\lambda\sigma}
\end{displaymath}

\noindent (Note that in the absence of spin-orbit coupling, there
is no spin mixing in these eigenstates and, since the matrix
elements of $H_{TBHF}$ are real, it is always possible to find a
set of eigenvectors whose components are real and denoted as
$c^{0n}_{i\lambda\sigma}$ in the following). The average values of
the three spin components are given by:

\begin{displaymath}
<{\bf S}>=\sum_{n\ \text{occ}}\bra{\psi_n}\frac{\bm{\sigma}}{2}\ket{\psi_n}
\end{displaymath}

\noindent in a non orthogonal orbital basis set, we obtain:

\begin{eqnarray}
<S_{x''}>&=&\text{Re}\sum_{ \substack{i\lambda,j\mu \\ n\text{ occ}}}c_{i\lambda\uparrow}^{n*}c_{j\mu\downarrow}^n
\mathcal{S}_{ij}^{\lambda\mu}  \nonumber \\
<S_{y''}>&=&\text{Im}\sum_{\substack{i\lambda,j\mu \\ n\text{ occ}}}c_{i\lambda\uparrow}^{n*}c_{j\mu\downarrow}^n
\mathcal{S}_{ij}^{\lambda\mu} \nonumber \\
<S_{z''}>&=&\frac{1}{2}\sum_{\substack{i\lambda,j\mu \sigma \\ n\text{ occ}}}
\sigma c_{i\lambda\sigma}^{n*}c_{j\mu\sigma}^n
\mathcal{S}_{ij}^{\lambda\mu} \label{eq:spintot}
\end{eqnarray}

\noindent i.e., in the absence of spin-orbit coupling
$<S_{x''}>=<S_{y''}>=0$.

For a system with full translational symmetry and a single atom
per unit cell, the Bloch theorem yields ($\sigma=\uparrow$ or $\downarrow$):

\begin{equation}
c_{i\lambda\sigma}^n=\frac{1}{\sqrt{N_{\text{at}}}}\exp(i{\bf k}.{\bf
R}_i)c_{\lambda\sigma}^{\alpha}({\bf k}) \label{eq:bloch}
\end{equation}

\noindent since each eigenstate $n$ is labelled by a band index
($\alpha=1,9$) and a wave vector ${\bf k}$. $N_{\text{at}}$ is the number of
atoms. The spin components are the same on all sites $i$ and are
given by:

\begin{eqnarray}
<S_{x''}>&=&\text{Re} \sum_{\substack{\lambda\mu \\ (\alpha,{\bf k}) \text{ occ}}}c_{\lambda\uparrow}^{\alpha*}({\bf
k})c_{\mu\downarrow}^{\alpha}({\bf k})
\mathcal{S}_{\lambda\mu}({\bf k}) \nonumber  \\
<S_{y''}>&=&\text{Im}\sum_{\substack{\lambda\mu \\ (\alpha,{\bf k}) \text{ occ}}}c_{\lambda\uparrow}^{\alpha*}({\bf
k})c_{\mu\downarrow}^{\alpha}({\bf k})
\mathcal{S}_{\lambda\mu}({\bf k})  \label{eq:spintot_periodic}\\
<S_{z''}>&=&\frac{1}{2}\sum_{\substack{\lambda\mu \sigma \\ (\alpha,{\bf k}) \text{ occ}}}
\sigma c_{\lambda\sigma}^{\alpha*}({\bf
k})c_{\mu\sigma}^{\alpha}({\bf k})
\mathcal{S}_{\lambda\mu}({\bf k})  \nonumber
\end{eqnarray}

\noindent with $\mathcal{S}_{\lambda\mu}({\bf
k})=N_{\text{at}}^{-1}\sum_{ij}\exp(i{\bf k}.({\bf R}_j-{\bf
R}_i))\mathcal{S}_{ij}^{\lambda\mu}$.

When all atoms are not
geometrically equivalent, we can define a spin on site $i$ by
identifying in Eqs.\ref{eq:spintot} all the terms involving this
site and, similarly to what is done to define Mulliken charges,
the overlap cross terms (i.e., those in which only one of the site
indices is equal to $i$) are multiplied by a factor $1/2$ to avoid
a double counting of these terms. This condition ensures that
these ``local'' spins are real and that their sum is equal to the
total spin. For example in  a periodic system with several atoms
per unit cell the local spin on each atom in the cell are given
by equations similar to equation \ref{eq:spintot_periodic} with an
additional index, labelling the atom in the cell. For instance in
a slab the local spin moment $<S_{a z"}>$ on layer $a$ is given by:

\begin{equation}
<S_{a z"}>=\frac{1}{4}\Bigg(\sum_{\substack{b\lambda\mu \sigma \\ (\alpha,{\bf k_{\pp}}) \text{ occ}}}
\sigma \big( c_{a\lambda\sigma}^{\alpha*}({\bf k_{\pp}})c_{b\mu\sigma}^{\alpha}({\bf k_{\pp}})\mathcal{S}_{\lambda\mu}^{ab}({\bf k_{\pp}})+
             c_{b\mu\sigma}^{\alpha*}({\bf k_{\pp}})c_{a\lambda\sigma}^{\alpha}({\bf k_{\pp}})\mathcal{S}_{\mu\lambda}^{ba}({\bf k_{\pp}})\big) \Bigg)
\end{equation}

\noindent
with $\mathcal{S}_{\lambda\mu}^{ab}({\bf k_{\pp}})=N_{\pp\text{at}}^{-1}
\sum_{\tilde{\text{\i}}\tilde{\text{\j}}}\exp(i{\bf k_{\pp}}.({\bf R}_{\tilde{\text{\j}}}-{\bf R}_{\tilde{\text{\i}}}))
\mathcal{S}_{\tilde{\text{\i}}a \tilde{\text{\j}}b  }^{\lambda\mu}$. $N_{\pp\text{at}}$ is the number of atoms in each layer of the
slab and ${\bf k_{\pp}}$ the wave vector parallel to the surface, each atom being now labelled by
a cell index, $\tilde{\text{\i}}$ or $\tilde{\text{\j}}$, and a layer index, $a$ or $b$. Corresponding
changes must be made for the two other components of $\bf S$.
Finally let us recall that the spin magnetic moment
${\bf M}$ is related to the spin ${\bf S}$ by $< {\bf M}>=-2<{\bf S}>$
(in Bohr magnetons $\mu_B$).

\subsubsection{Orbital moment.}

Up to now the orbital moment in the TB approximation has always
been calculated by assuming an orthogonal basis set of atomic
orbitals and only its $z''$ component was determined. In these
conditions, the component of the local orbital moment on site $i$
in this direction is usually written as \cite{Guirado03}:

\begin{equation}
<L_{iz''}>=\sum_{lm\sigma}m\int^{E_F}_{-\infty}\rho_{ilm\sigma}(E)dE
\label{eq:Lric}
\end{equation}

\noindent where $\rho_{ilm\sigma}(E)$ is the local density of
states at site $i$ projected on the atomic orbitals
$\ket{ilm}=\mathcal{R}_l(r'')Y_{lm}(\theta'',\varphi'')$ and spin function
$\sigma$, the variable $r'', \theta'', \varphi''$ being spherical coordinates
relative to the spin framework, i.e.:

\begin{equation}
\rho_{ilm\sigma}(E)=\sum_{\substack{lm \\ n}}\braket{\psi_n}{ilm\sigma}\braket{ilm\sigma}{\psi_n}\delta(E-E_n).
\end{equation}

Thus

\begin{equation}
<L_{iz''}>=\sum_{\substack{lm\sigma \\n \text{ occ}}}
\braket{\psi_n}{ilm\sigma}m\braket{ilm\sigma}{\psi_n} \label{eq:Liz}
\end{equation}

This defines the operator $L_{iz''}$ which is diagonal in the
$\ket{ilm\sigma}$ basis. Eq.\ref{eq:Liz} can be generalized for the
two other components of the orbital moment by noting that the
corresponding operators are not diagonal in this basis. This
gives:

\begin{equation}
<{\bf L}^{''}_i>=\sum_{\substack{lm,l'm'\sigma \\ n \text{ occ}}}
\braket{\psi_n}{ilm\sigma}
[{\bf L}^{''}_i]_{lm,l'm'} \braket{il'm'\sigma}{\psi_n} \label{eq:Lvec}
\end{equation}

\noindent with ${\bf L}^{''}_i=(L_{ix''}, L_{iy''}, L_{iz''})$.
Finally in the basis of real orbitals $\ket{i\lambda\sigma}$ defined
in the crystal frame, we have:

\begin{equation}
<{\bf L}^{''}_i>=\sum_{ \substack{\lambda\mu\sigma \\ n\text{ occ}}}
c^{n*}_{i\lambda\sigma}[{\bf L}^{''}_i]_{\lambda\mu}c^n_{i\mu\sigma} \label{eq:Lseci}
\end{equation}

The operators ${\bf L}^{''}_i$ can be expressed as a function of
the three operators $L_{ix}, L_{iy}, L_{iz}$ projecting the
orbital moment on the crystal axes, i.e.:

\begin{eqnarray}
L_{ix''} &=& \cos\theta\cos\varphi\ L_{ix}+\cos\theta\sin\varphi\
L_{iy}-\sin\theta\ L_{iz} \nonumber \\
L_{iy''} &=& -\sin\varphi\ L_{ix}+\cos\varphi\ L_{iy} \nonumber \\
L_{iz''} &=& \sin\theta\cos\varphi\ L_{ix}+\sin\theta\sin\varphi\
L_{iy}+\cos\theta\ L_{iz} \label{eq:L}
\end{eqnarray}

\noindent and the matrix elements of ${\bf L}_i$ between two
atomic orbitals $\lambda$ and $\mu$ centered on atom $i$ defined
with respect to the crystal axes are easily calculated (see
Appendix A). These matrix elements are either vanishing or
imaginary, thus $[{\bf L}_i]_{\lambda\mu}=-[{\bf
L}_i]_{\mu\lambda}$. In the absence of spin-orbit coupling, as
stated above, the coefficients $c^{0n}_{i\lambda\sigma}$ are real
and the orbital moment vanishes, i.e., $<{\bf L}_i^{''}>=0$.

Eq.\ref{eq:Lseci} can be generalized to take overlap into account
(see Appendix B). This yields

\begin{equation}
<{\bf L}^{''}_i>=\text{Re}
\sum_{ \substack{\lambda\mu j\nu\sigma \\ n \text{ occ}}}c^{n*}_{i\lambda\sigma}[{\bf
L}^{''}_i]_{\lambda\mu}\mathcal{S}^{\mu\nu}_{ij}c^n_{j\nu\sigma}
\end{equation}

\noindent and, for a system with a full translational symmetry and a single atom per unit cell,
this gives using Eq.\ref{eq:bloch}:

\begin{equation}
<{\bf L}^{''}_i>=\text{Re}
\sum_{ \substack{ \lambda\mu\nu\sigma \\ \alpha{\bf k} \text{ occ}} }c^{\alpha*}_{\lambda\sigma}({\bf k})
[{\bf L}^{''}_i]_{\lambda\mu}\mathcal{S}_{\mu\nu}({\bf k})
c^{\alpha}_{\nu\sigma}({\bf k}) \label{eq:lk}
\end{equation}

This latter equation can be easily generalized to periodic systems with
several atoms per unit cell, in the same way as for the spin moment.
Finally, let us note that in the presence of spin-orbit coupling,
the direction of the total magnetization ($-<{\bf L}+2{\bf S}>$)
may not strictly be parallel to the spin quantization axis $z''$.
However $H_{\text{so}}$ being a small perturbation in Fe, we will often
denote the spin quantization axis as the magnetization direction
in the following.

\subsection{Magnetocrystalline anisotropy and orbital moment from
perturbation theory}

We have seen in section 2.2 that spin-orbit effects can be limited to
$d$ orbitals. Furthermore the overlaps between these orbitals are
close to zero and the spin-orbit coupling is a weak perturbation
since $\xi$ is much smaller than the Fe $d$ bandwidth.
Consequently spin-orbit coupling effects can be understood using a
simple perturbation theory with a basis set of orthogonal $d$
orbitals \cite{Bruno89th,Bruno89,vanderLaan98}.

Let us consider the perturbation of the total energy due to
$H_{\text{so}}$. Since the matrix elements of $H_{\text{so}}$ are a function of
$\theta$  and $\varphi$, this introduces an angular dependence of
this perturbation which is known as the magnetocrystalline
anisotropy. The first order term can be written:

\begin{equation}
\Delta E^{(1)}=\sum_{n\sigma \text{ occ}} \bra{n\sigma}H_{\text{so}}\ket{n\sigma}
\end{equation}

\noindent where $\ket{n\sigma}$ is an unperturbed state of energy
$E^0_{n\sigma}$, i.e.,

\begin{equation}
\ket{n\sigma}=\sum_{i\lambda}c^{0n}_{i\lambda\sigma}\ket{i\lambda\sigma}
\label{eq:ups}
\end{equation}

\noindent Thus:

\begin{equation}
\Delta E^{(1)}=\xi \sum_{\lambda\mu}\bra{\bar{\lambda}\sigma} {\bf
L}.{\bf S}\ket{\bar{\mu}\sigma} \sum_{\substack{ i \\ n \sigma \text{ occ}}}
c^{0n}_{i\lambda\sigma}c^{0n}_{i\mu\sigma}
\end{equation}

\noindent It is easily seen that $\Delta E^{(1)}$ vanishes since
for each spin the $(5\times 5)$ matrix $ \bra{\bar{\lambda}\sigma}{\bf
L}.{\bf S}\ket{\bar{\mu}\sigma}$ is imaginary (see appendix A).

The second order perturbation of the total energy is given by:

\begin{equation}
\Delta E^{(2)}=-\sum_{\substack{ n\sigma \text{ occ} \\ n'\sigma'\text{ unocc}}}
\frac{|\bra{n\sigma}H_{so}\ket{n'\sigma'}|^2}{E^0_{n'\sigma'}-E^0_{n\sigma}}.
\end{equation}

\noindent This yields:

\begin{equation}
\Delta
E^{(2)}=-\xi^2\sum_{\lambda\mu\lambda'\mu'}\sum_{\sigma\sigma'}
\bra{\bar{\lambda}\sigma}{\bf L}.{\bf S}\ket{\bar{\mu}\sigma'}
\bra{\bar{\mu'}\sigma'}{\bf L}.{\bf S}\ket{\bar{\lambda'}\sigma}
\sum_{ij}I_{ij}(\lambda,\lambda',\mu',\mu,\sigma,\sigma')
\end{equation}

\noindent with:

\begin{equation}
I_{ij}(\lambda,\lambda',\mu',\mu,\sigma,\sigma')=\int^{E_F}_{-\infty}dE
\int_{E_F}^{\infty}dE'\frac{\rho^{0\lambda\lambda'}_{ij\sigma}(E)\rho^{0\mu'\mu}_{ji\sigma'}(E')}
{E'-E}
\end{equation}

\noindent and:

\begin{equation}
\rho^{0\lambda\lambda'}_{ij\sigma}(E)=\sum_nc^{0n}_{i\lambda\sigma}c^{0n}_{j\lambda'\sigma}\delta(E-E^0_{n\sigma})
\end{equation}

\noindent By using the relations between the matrix elements of
${\bf L}.{\bf S}$ shown by Bruno \cite{Bruno89th,Bruno89} and recalled in Appendix A, $\Delta
E^{(2)}$ can be rewritten as:

\begin{eqnarray}
\Delta E^{(2)}&=& \text{isotropic term}
 \nonumber \\ 
&-&\xi^2\sum_{\lambda\mu\lambda'\mu'}
\bra{ \bar{\lambda}\uparrow}{\bf L}.{\bf S}\ket{\bar{\mu}\uparrow}
\bra{ \bar{\mu'}\uparrow}   {\bf L}.{\bf S}\ket{ \bar{\lambda'}\uparrow}
\sum_{ij,\sigma\sigma'}\sigma\sigma'I_{ij}
(\lambda,\lambda',\mu',\mu,\sigma,\sigma') \nonumber \\
&=&\sum_i\Delta E^{(2)}_i \label{eq:de2}
\end{eqnarray}

\noindent where $\Delta E^{(2)}_i$ is the contribution of atom $i$
to the perturbation energy.

In the case of a system with full translational symmetry and a
single atom per unit cell and using Eq.\ref{eq:bloch},
Eq.\ref{eq:de2} can be transformed into \cite{Bruno89th,Bruno89}:

\begin{eqnarray}
\lefteqn{ \Delta E^{(2)}=\text{isotropic term}
-\xi^2\sum_{\lambda\mu\lambda'\mu'}
\bra{\bar{\lambda}\uparrow}{\bf L}.{\bf S} \ket{\bar{\mu}\uparrow}
\bra{\bar{\mu'}\uparrow}   {\bf L}.{\bf S} \ket{\bar{\lambda'}\uparrow} } \nonumber \\
& &  \times \sum_{\bf k}\int^{E_F}_{-\infty}dE
\int_{E_F}^{\infty}dE'\frac{\mathcal{M}_{\lambda\lambda'}({\bf
k},E)\mathcal{M}_{\mu'\mu}({\bf k},E')} {E'-E} {}
\end{eqnarray}

\noindent with:

\begin{equation}
\mathcal{M}_{\lambda\lambda'}({\bf
k},E)=\mathcal{N}_{\lambda\lambda'\uparrow}({\bf
k},E)-\mathcal{N}_{\lambda\lambda'\downarrow}({\bf k},E)
\end{equation}

\noindent and:

\begin{equation}
\mathcal{N}_{\lambda\lambda'\sigma}({\bf
k},E)=\sum_{\alpha}c^{0\alpha*}_{\lambda\sigma}({\bf
k})c^{0\alpha}_{\lambda'\sigma}({\bf
k})\delta(E-E^0_{\alpha\sigma}({\bf k}))
\end{equation}

\noindent the superscript $0$ refers to the unperturbed state as
above and $E^0_{\alpha\sigma}({\bf k})$ are the unperturbed
eigenenergies.These equations can be generalized to a periodic system
with several atoms per unit cell \cite{Cinal94}.

Let us now consider the projection of the orbital moment on the
spin framework axes. From Eq.\ref{eq:Lseci} it can be seen that
the operators associated with these projections at a given site
$i$ can be written in an orthogonal basis set:

\begin{equation}
{\bf L}^{''}_i=\sum_{\lambda\mu\sigma}\ket{i\lambda\sigma}[{\bf
L}^{''}_i]_{\lambda\mu}\bra{i\mu\sigma}
\label{eq:OMO}
\end{equation}

\noindent Within perturbation theory, we have:

\begin{equation}
<{\bf L}^{''}_i>=-\sum_{\substack{n\sigma \text{ occ} \\ n'\sigma' \text{ unocc}}}
\frac{\bra{n\sigma}{\bf L}^{''}_i\ket{n'\sigma'}\bra{n'\sigma'}H_{\text{so}}\ket{n\sigma}}
{E^0_{n'\sigma'}-E^0_{n\sigma}}+ \text{ c.c.}
\label{eq:lp}
\end{equation}

\noindent By substituting Eqs.\ref{eq:OMO} and \ref{eq:ups} for
${\bf L}^{''}_i$ and $\ket{n\sigma}$, respectively, into the preceding
equation, we get:

\begin{equation}
<{\bf
L}^{''}_i>=-2\xi\sum_{\lambda\mu\lambda'\mu'}\sum_{\sigma}
\bra{\bar{\lambda}\sigma}{\bf L}^{''}_i\ket{\bar{\mu}\sigma}
\bra{\bar{\mu'}\sigma}   {\bf L}.{\bf S}\ket{\bar{\lambda'}\sigma}
\sum_{j}I_{ij}(\lambda,\lambda',\mu',\mu,\sigma,\sigma)
\label{eq:aom}
\end{equation}

\noindent the factor 2 in Eq. \ref{eq:aom} accounts for the complex conjugate in Eq.\ref{eq:lp}
since the matrix elements of ${\bf L}^{''}_i$ and ${\bf L}.{\bf S}$ for parallel spins are
imaginary and all $I_{ij}$ are real. For a system we full translational symmetry and a single
atom per unit cell this equation becomes \cite{Bruno89}:

\begin{eqnarray}
\lefteqn{<{\bf L}^{''}_i>=
-2\xi\sum_{\lambda\mu\lambda'\mu'}\sum_{\sigma}
\bra{\bar{\lambda}\sigma}{\bf L}^{''}_i\ket{\bar{\mu}\sigma}
\bra{\bar{\mu'}\sigma}   {\bf L}.{\bf S}\ket{\bar{\lambda'}\sigma} }  \\
& &  \times \sum_{\bf k}\int^{E_F}_{-\infty}dE
\int_{E_F}^{\infty}dE'\frac{\mathcal{N}_{\lambda\lambda'\sigma}({\bf
k},E)\mathcal{N}_{\mu'\mu\sigma}({\bf k},E')} {E'-E} \nonumber
\label{eq:LNN}
\end{eqnarray}

Furthermore noting that $[L_{iz''}]_{\lambda\mu}=
2\sigma\bra{\bar{\lambda}\sigma}{\bf L}.{\bf S}\ket{\bar{\mu}\sigma}$,
Eq.\ref{eq:aom} for $<L_{iz''}>$ can be transformed into:

\begin{equation}
<
L_{iz''}>=
-4\xi\sum_{\lambda\mu\lambda'\mu'}\sum_{\sigma}
\sigma \bra{\bar{\lambda}\uparrow}{\bf L}.{\bf S}\ket{\bar{\mu}\uparrow}
       \bra{\bar{\mu'}\uparrow}   {\bf L}.{\bf S}\ket{\bar{\lambda'}\uparrow}
       \sum_{j}I_{ij}(\lambda,\lambda',\mu',\mu,\sigma,\sigma)
\label{eq:Lizs}
\end{equation}

\noindent For a system with full translational symmetry and one
atom per unit cell, this yields \cite{Bruno89}

\begin{eqnarray}
\lefteqn{<L_{iz''}>=
-2\xi\sum_{\lambda\mu\lambda'\mu'}
\bra{\bar{\lambda}\uparrow}{\bf L}.{\bf S}\ket{\bar{\mu}\uparrow}
\bra{\bar{\mu'}\uparrow}{\bf L}.{\bf S}\ket{\bar{\lambda'}\uparrow} }  \\
& & \times\sum_{\bf k}\int^{E_F}_{-\infty}dE
\int_{E_F}^{\infty}
dE'\frac{\mathcal{N}_{\lambda\lambda'}({\bf k},E)\mathcal{M}_{\mu'\mu}({\bf k},E')+
         \mathcal{M}_{\lambda\lambda'}({\bf k},E)\mathcal{N}_{\mu'\mu}({\bf k},E')} {E'-E} \nonumber
\end{eqnarray}

\noindent in which
$\mathcal{N}_{\lambda\lambda'}({\bf k},E)=\sum_{\sigma}\mathcal{N}_{\lambda\lambda'\sigma}({\bf k},E)$.
The generalization of this equation to systems with several atoms per unit cell
is straightforward.

It can be seen from Eqs.\ref{eq:de2} and \ref{eq:Lizs} that $<
L_{iz''}>$ and the anisotropic part of $\Delta E^{(2)}_i$ are both
given by quadratic functions of the direction cosines of the spin
quantization axis relative to the crystal ones since the involved
matrix elements of ${\bf L}.{\bf S}$ are all proportional to one
of these direction cosines (see Appendix A). These two functions
present some similarity, but spin-flip excitations contribute to
$\Delta E^{(2)}_i$ but not to $< L_{iz''}>$. However, if the
exchange splitting is large enough compared to the $d$ bandwidth,
the spin up band is completely filled and the contribution of
spin-flip excitations to $\Delta E^{(2)}_i$ is negligible due to
the large value of the energy denominator. In this condition, for
each site $i$, the anisotropy of $\Delta E^{(2)}_i$ and $<
L_{iz''}>$ are proportional:

\begin{equation}
\Delta E^{(2)}_i(\theta,\varphi)-\Delta E^{(2)}_i(0,0)=-\frac{\xi}{4}(<L_{iz''}(\theta,\varphi)>-< L_{iz''}(0,0)>)
\label{eq:rap_ani}
\end{equation}

\noindent Note that this relation was already derived by Bruno
\cite{Bruno89th} for fcc monolayers with a single atom per unit
cell. Finally let us point out that for bulk cubic crystals with a single atom per unit cell
both $\Delta E^{(2)}$ and $< L_{iz''}>$ are isotropic at the
orders of perturbation considered.

\subsection{The ab-initio method}

For the sake of comparison we have also performed spin-polarized ab-initio calculations
based on the Density Functional Theory (DFT)
using the PWscf code of $\nu$-ESPRESSO package\cite{espresso} with ultrasoft pseudopotentials including non-linear
core corrections. The calculations without spin-orbit coupling
have been carried out within the GGA and the Perdew-Wang exchange-correlation parametrization,
while the one including spin-orbit coupling
have been performed within the LDA and the Perdew-Zunger exchange-correlation
parametrization. The plane wave kinetic-energy cut-off was taken equal to
35Ry for the wavefunctions and 250Ry for the charge density and potential,
which ensures a very good energy precision.

\subsection{Computational details}

When dealing with magnetic properties, and in particular magnetic anisotropy,
the convergence of the total energy with respect to the number of
$k$-points has to be checked carefully. For all calculations involving magnetic
anisotropy we checked that our results did not change by more than a few
hundredth of meV (at most 0.1meV in the worst case).
In the case of PWscf calculations the use of plane
waves imposes a periodically repeated geometry and one must also
avoid as much as possible electronic interactions by using large unit cells.
The monatomic wires were separated by 30a.u., but for surfaces, we have been less
demanding since we only calculated the magnetic moment
and therefore the slabs were separated by approximately 17a.u.

\section{Bulk magnetism of bcc and fcc iron}

\subsection{Determination of the Stoner parameter I$_{dd'}$ from the magnetic
transition in bcc iron}

\begin{figure}[!fht]
\begin{center}
\includegraphics*[scale=0.4,angle=0]{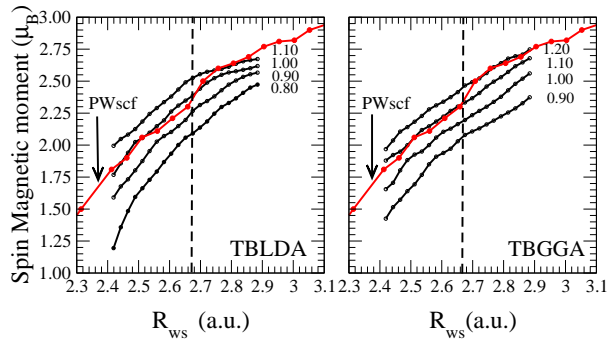}
\end{center}
\caption{Variation of the absolute value of the spin magnetic moment (per atom) of bcc Fe as a function of
 the Wigner-Seitz radius $R_{WS}$ for the values of the Stoner parameter $I_{dd'}$ (in eV) given
 on each curve. Left and right panels correspond
 to TBLDA and TBGGA calculations, respectively, compared to PWscf calculations in GGA.
 The dashed vertical line gives the experimental Wigner-Seitz radius at equilibrium.}
\label{fig:M_d_Febcc}
\end{figure}

In our TBHF model, the magnetism is entirely governed by the value
of the Stoner parameter $I_{dd'}$. It is well known that, in
unsaturated magnetic materials like Fe, the magnetic moment is very
sensitive to the precise value of the equilibrium interatomic
distance. As a general trend, an expansion of the bulk lattice
parameter leads to narrower (thus higher) density of states, which
usually plays in favor of magnetism: it increases the
magnetization in magnetic materials or it can trigger a magnetic
transition in non-magnetic materials. Therefore a straightforward
way to determine $I_{dd'}$ is  to study the evolution of the
magnetic moment as a function of the lattice parameter. In Fig.
\ref{fig:M_d_Febcc} the result of a series of TBLDA and TBGGA
calculations on bulk bcc iron is shown for various values of the
Stoner parameter. As expected the magnetic moment increases when
the lattice is expanded but also when the Stoner parameter is
increased. With $I_{dd'}=1eV$ (TBLDA) and $I_{dd'}=1.10eV$ (TBGGA)
we have been able to  reproduce closely the results of PWscf
calculations in a range of Wigner-Seitz radii ($R_{WS}$) around
equilibrium (the experimental bcc lattice parameter, 2.87\AA,
corresponds to $R_{WS}=2.67$ a.u.). In the following we will keep
these values fixed and neglect any variation of these parameters
with the local atomic environment. Let us however note that at very
large lattice spacings the spin moment saturates (not seen in
Fig.\ref{fig:M_d_Febcc}) at an ``atomic'' value of 2 $\mu_B$ for
TBLDA and  4 $\mu_B$ for TBGGA. These two limits  correspond to
the different atomic configurations $3d^8 4s^0$ and $3d^64s^2$
found in TBLDA and TBGGA, respectively, for a free Fe atom.
Therefore the TBLDA  gives a wrong atomic configuration which will
have some consequences on the surface magnetism. Let us finally
mention that in ab-initio calculations LDA and GGA yield very
similar results as far as the magnetic moment is concerned.

\subsection{Fcc iron}

The ground state phase of iron in normal temperature and pressure conditions is ferromagnetic (FM) bcc, at
higher temperatures,  the fcc phase is stabilized but in a NM configuration.
However it has been shown experimentally that thin films of iron can be stabilized in
an fcc structure\cite{Macedo88,Torija05}. This experimental work also showed the existence of various magnetic phases.
It has also been known for a long time from theoretical works 
\cite{Kubler81,Moruzzi86,Herper99,Knopfle00, Neumann04} 
that fcc iron has a much more complicated magnetic structure than bcc iron. In the following, we present a study of the
magnetic properties of fcc Fe with TBGGA parameters. This will provide us with a first
check of our model.

\subsubsection{Magnetic transition in fcc iron.}

\begin{figure}[!fht]
\begin{center}
\includegraphics*[scale=0.4,angle=0]{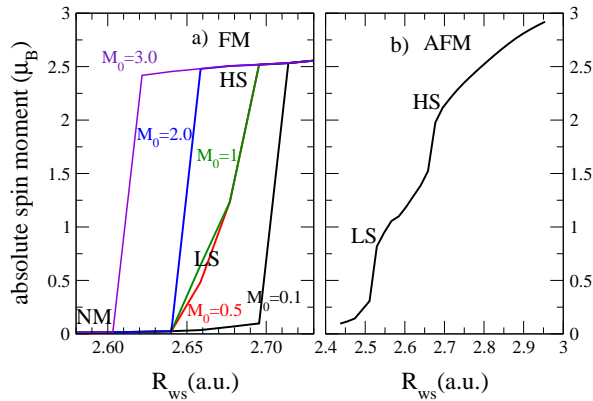}
\end{center}
\caption{Variation of the absolute value of the magnetic moment (per atom) for the ferromagnetic (FM) and
antiferromagnetic (AFM) states of fcc Fe as a function of the Wigner-Seitz radius $R_{WS}$,  for various input
magnetic moments $M_{0}$ in $\mu_B$ obtained with TBGGA parameters. LS and HS denote
respectively low and high spin states. }
\label{fig:M_d_Fefcc}
\end{figure}

We have carried out a series of calculations on the fcc phase of
iron. Fig. \ref{fig:M_d_Fefcc} is similar to Fig.
\ref{fig:M_d_Febcc} but for the FM (a) and antiferromagnetic (AFM)(b)
bulk fcc phases, the latter
corresponding to a stacking of $(001)$ planes in which spins of
adjacent layers are opposite. In the FM case (Fig.
\ref{fig:M_d_Fefcc}a), the curves of appearance of a magnetic
moment show a strong dependence on the magnetic moment $M_{0}$
chosen as input to begin the self-consistency iterations (a
similar behavior is also found with the PWscf code). For large
$M_{0}$ an abrupt transition from a NM configuration to a High
Spin (HS) state occurs, for instance at $R_{WS}\simeq 2.6$a.u.
when $M_0=3\mu_B$. For a small value of $M_{0}$ a similar
transition appears but at a much larger volume ($R_{WS}\simeq
2.7$a.u), while for intermediate $M_{0}$ this transition is less
steep. The magnetic transition is much less abrupt in the
antiferromagnetic phase (Fig.\ref{fig:M_d_Fefcc}b)  where two
steps are observed: first, from a NM state to a low spin state
(LS) and, second, from the LS state to the HS state. Note that a
similar dependence on the input magnetic moment also exists in the AFM state
(not shown on the graph).

\subsubsection{Low and High spin ferromagnetic states: fixed spin moment calculations.}

\begin{figure}[!fht]
\begin{center}
\includegraphics*[scale=0.4,angle=0]{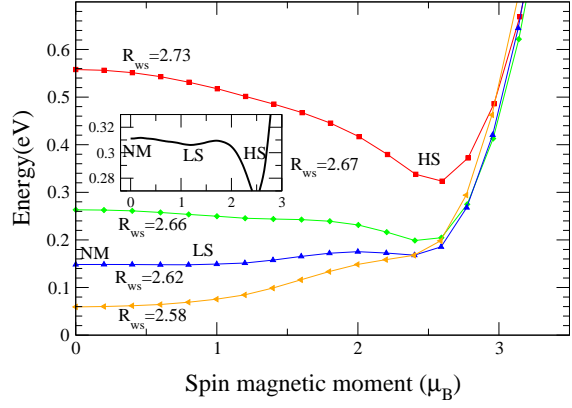}
\end{center}
\caption{Variation of the TBGGA total energy per atom  with the
magnetic moment (fixed spin moment calculation for the
ferromagnetic state) for fcc Fe at several Wigner-Seitz radii
$R_{WS}$ in a.u.. Note the presence of stable (or metastable) non
magnetic (NM), Low spin (LS)  and High spin (HS) states which is
clearly seen in the inset. The zero of energy is arbitrary.}
\label{fig:E_M_Fefcc}
\end{figure}

The strong dependence on the input magnetic moment suggests the existence of metastable magnetic
solutions. Therefore we have carried out TBGGA calculations using a fixed spin moment
procedure for a series of Wigner-Seitz radii corresponding to the region of the magnetic transition.
The behavior of the total energy as a function of the total moment M
(Fig. \ref{fig:E_M_Fefcc}) reveals the existence of several
local minima. In particular the curve at $R_{WS}=2.67$a.u. exhibits three minima (inset of Fig.\ref{fig:E_M_Fefcc}):
one at $M=0$, one around $M=1.2\mu_B$ and one at $M=2.5\mu_B$,
corresponding  to the NM, LS and HS states, respectively. This complex energy behavior is in
agreement with the results obtained by Moruzzi et al. \cite{Moruzzi89}  who showed for the first time the existence of
three phases. It is clear that depending on the value
of the input moment, the iteration loop will converge towards one of the three
self-consistent (stable or metastable) magnetic states.

\subsection{Relative phase stability: comparison between TBLDA and TBGGA models.}

\begin{figure}[!fht]
\begin{center}
\includegraphics*[scale=0.4,angle=0]{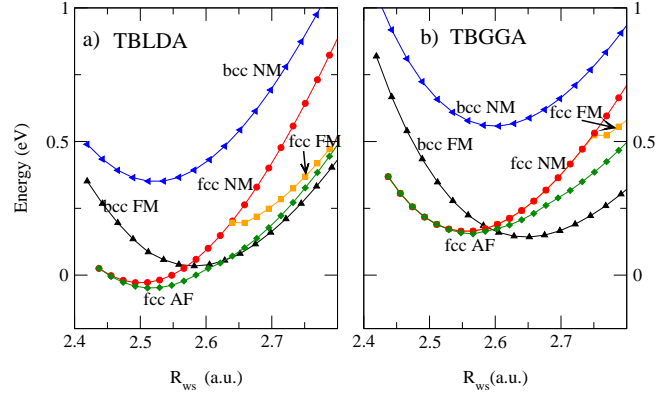}
\end{center}
\caption{Total energy per atom as a function of the Wigner Seitz radius $R_{WS}$ for ferromagnetic (FM), antiferromagnetic
 (AFM) and non magnetic (NM) states of bcc and fcc iron. Panels a) and b) correspond to calculations
 performed with  TBLDA and  TBGGA, respectively. The zero of energy is arbitrary but is the same
 for all the curves in each panel.}
\label{fig:E_d_TBLDA_TBGGA}
\end{figure}

It is well known that DFT in the LDA predicts
that, at low temperature, the fcc AFM phase is the most stable one contrary to experiments \cite{Wang85}.
However the right phase stability is recovered with the GGA.
It is therefore interesting to investigate the ground state  properties of bulk iron within
our TB model. The results are presented in Fig. \ref{fig:E_d_TBLDA_TBGGA},
it is found that TBLDA, similarly to DFT-LDA, gives the AFM state
of fcc iron as the most stable phase, while with TBGGA the  FM bcc phase is found to be the
ground state.  These results are in perfect agreement with ab-initio findings, showing the ability of
our model to reproduce rather complex magnetic behaviors.

\subsection{Influence of spin-orbit coupling}

The results presented above have been obtained without including
spin-orbit coupling effects. The value of the spin-orbit coupling
parameter $\xi$ can be deduced from a comparison of the NM bcc band
structure along a high symmetry direction of the Brillouin zone,
for instance $\Gamma H$, obtained with our model and with the
PWscf code. Indeed, the effect of spin-orbit coupling is to remove
the degeneracy of degenerate levels when a matrix element of
$H_{\text{so}}$ exists between the corresponding eigenstates (see
Appendix A). For instance it is easily seen that the six-fold degenerate
level $\Gamma^0_{25'}$, corresponding to $t_{2g}$ spin-orbitals, are
coupled by some matrix elements of $H_{\text{so}}$. Using the perturbation
theory for degenerate levels, it is easily found that this level
splits into a four-fold degenerate level at $\Gamma^0_{25'}-\xi/2$
and a doubly degenerate level at $\Gamma^0_{25'}+\xi $. From this
splitting calculated with the PWscf code, we obtain $\xi=0.06eV$.
We have verified that with this value, our model is able to
reproduce perfectly the spin-orbit coupling effects along $\Gamma H$.

As seen in section 2, spin-orbit coupling is at the origin of the
magneto-crystalline anisotropy. However, since this anisotropy is
of fourth order in $\xi$, typical values for bulk materials are
very small and of the order of $10^{-5}-10^{-6}$ eV per atom for
Fe, Co or Ni which makes the calculation of this anisotropy almost
impossible, since it is beyond the accuracy of electronic
structure methods. On the contrary, reliable values of the orbital
moment, which is isotropic in the bulk to first order in
perturbation, can be derived from Eq.\ref{eq:lk}. With our model
and TBGGA parameters we find $<L_z>=0.07\mu_B$ in good agreement
with experiments ($0.08\mu_B$ \cite{Landolt86}).

As a conclusion, the TB results presented above (section 3.2 and 3.3) are
in perfect agreement with DFT calculations, showing the ability of
our model to reproduce rather complex magnetic behaviors.

\section{$(110)$ and $(001)$ surfaces of iron}

We have then applied our TBGGA model to the study of the $(001)$ and $(110)$
surfaces of bcc iron. At the surface some atoms have a reduced
coordination and therefore charge transfers as large as some
tenths of electron are found in our model if the atomic levels
$\varepsilon_s$, $\varepsilon_p$, $\varepsilon_d$ in Eq.\ref{eq:levels}
are kept at the values given by the MP equations. However, it is
known that in metals, due to screening, the charge transfers are
expected to be at least one order of magnitude smaller. To avoid
unphysical charge transfers at surfaces the Hamiltonian
is corrected by adding a term depending on the charge transfer  $\delta N_i$
and of an average Coulomb integral $U$ which must be  large
enough ($U=5$eV) as shown in Eq.\ref{eq:Htot}.

\subsection{Band structure of the (110) surface}

\begin{figure}[!fht]
\begin{center}
\includegraphics*[scale=0.4,angle=0]{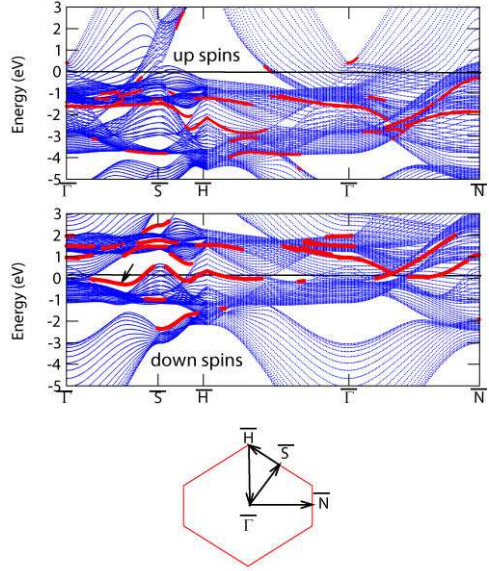}
\end{center}
\caption{TBGGA projected band structure for up (top) and down
(middle) spins of a 20-layer (110) slab of bcc Fe with the
experimental lattice parameter of 2.87\AA. The energy zero is the
Fermi level. Surface or resonant states (i.e., states with more
than 60\% of their total weight on the first two outer layers) are
represented in red and with thicker dots. A characteristic surface
state of minority spin is indicated by an arrow. A schematic
representation of the surface Brillouin zone and of the path in
the reciprocal space is shown at the bottom.}
\label{fig:surf110_bands}
\end{figure}

In a previous work on rhodium surfaces we showed that the charge
quasi-neutrality is crucial to obtain a good description of the
surface and resonant states\cite{Barreteau98}. Indeed these states are extremely
sensitive to the energy shift induced by the renormalization of
the intra-atomic terms of the TB Hamiltonian. Here we have carried
out a TBGGA projected band structure calculation for the $(110)$
surface of bcc Fe. The results are shown in Fig.
\ref{fig:surf110_bands}. It can be seen that the position and size
of pseudo-gaps in the band structure is significantly different
for up and down spins. In particular along the
$\bar{\Gamma}\bar{S}$ and $\bar{S}\bar{H}$ directions the pseudo-gaps
 are much larger in the minority spin band structure than in
the majority spin one. As a consequence there are more minority
spin than majority spin surface states. This is evidenced by the
presence of a sharp down spin surface state around the Fermi level
(indicated by an arrow in Fig. \ref{fig:surf110_bands}) which
disappears in the up spin band structure. These results are in
excellent agreement with previous ab-initio calculations
\cite{Kim01} in particular, for the position and dispersion
of the characteristic down spin surface state discussed above.

\subsection{Spin magnetic moments of Fe$(110)$ and Fe$(001)$ surfaces }

\begin{figure}[!fht]
\begin{center}
\includegraphics*[scale=0.4,angle=0]{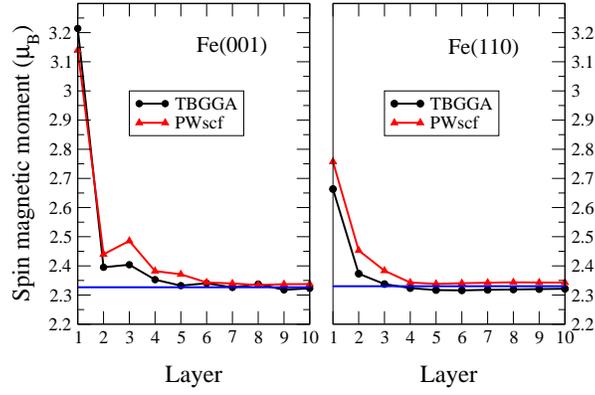}
\end{center}
\caption{Variation of the spin magnetic moment (per atom) on
successive atomic layers of $(001)$ and $(110)$ slabs (with 20
atomic layers) of bcc Fe obtained from the TBGGA model and PWscf
code with GGA. Layer 1 corresponds to the outermost layer and
layer 10 to a central layer. The value of the bulk magnetic moment
is indicated as a reference.} \label{fig:surf_spin_moment}
\end{figure}

It is well known that the lowering of coordination induces a
narrowing of the density of states which  usually enhances the
magnetic moment. Consequently, it is expected that open surfaces
should have larger surface magnetic moments than close-packed
ones. We have therefore carried out self consistent TBGGA and
PWscf GGA calculations for $(001)$ and $(110)$ surfaces. The
$(001)$ surface being more open than the $(110)$ one, since each
atom from their outermost layer looses 4 and 2 first nearest
neighbors, respectively, we expect larger surface magnetic moments
for the $(001)$  than for the $(110)$ surface. Fig.
\ref{fig:surf_spin_moment} shows that this general rule of thumb
is well obeyed. Actually two clear features are seen in Fig.
\ref{fig:surf_spin_moment}: the magnetic moment is more reinforced
on the $(001)$ than on the $(110)$ surface (+38\% and +14\%,
respectively, for the outermost layer compared to the bulk ).
Friedel type oscillations are present on the $(001)$ surface while
an almost monotonic decrease is obtained for the $(110)$ surface.
An excellent agreement is once again observed between PWscf and
TBGGA results, in particular the spin moment is almost saturated on the
outermost layer of the $(001)$ surface in both calculations.
Let us however note that the agreement is less
perfect within TBLDA (not shown), which can be attributed to the
wrong atomic configuration obtained in this model which
deteriorates the $spd$ charge distribution on the surface plane.

\subsection{Magneto-crystalline anisotropy}

For surfaces the magneto-crystalline anisotropy is usually one or
two orders of magnitude larger than in the bulk. Indeed, it is
well known that, contrary to the bulk, this anisotropy is of the
second order in $\xi$ at surfaces. Actually, we have seen in
section 2.4 that second order perturbation theory predicts that the
magneto-crystalline anisotropy is a quadratic function of the
direction cosines ($l=\sin\theta\cos\varphi,
m=\sin\theta\sin\varphi, n=\cos\theta$) of the spin quantization
axis relative to crystal axes. By imposing the symmetry properties
of the surface to this quadratic form the following laws are
easily derived:

\begin{equation}
\Delta E^{(2)}_i(\theta,\varphi)-\Delta
E^{(2)}_i(0,0)=K_1^{(001)}\sin^2\theta \label{eq:k001}
\end{equation}

\noindent for the $(001)$ surface with $x$ and $y$ crystal axes parallel to the
edges of the square two dimensional cell, and:

\begin{equation}
\Delta E^{(2)}_i(\theta,\varphi)-\Delta
E^{(2)}_i(0,0)=K_1^{(110)}\sin^2\theta+K_2^{(110)}\sin^2\theta\cos2\varphi
\label{eq:k110}
\end{equation}

\noindent on the $(110)$ surface. For this surface the crystal axes
are chosen as follows: the $z$ axis is perpendicular to the
surface and the $y$ one is parallel to the second nearest neighbor
direction in the surface.

\begin{figure}[!fht]
\begin{center}
\includegraphics*[scale=0.5,angle=0]{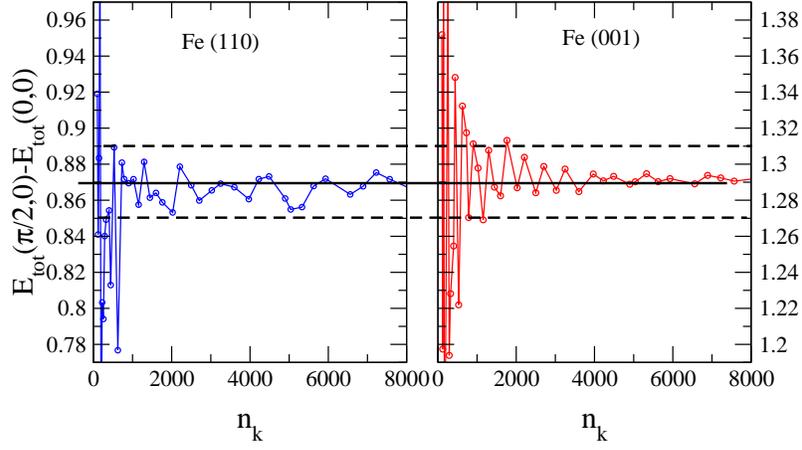}
\end{center}
\caption{Convergence of the magnetic anisotropy $E_{tot}(\pi/2,0)-E_{tot}(0,0)$ with respect
to the number of $k$ points $n_k$  in the first Brillouin zone, for $(110)$ and $(001)$  Fe bcc, unsupported
monolayers. The magnetic anisotropy is oscillating around its asymptotic value (full straight line) with an amplitude
 below $\pm 0.02$ meV when $n_k$ is larger than 1000.} \label{fig:conv_kpoints}
\end{figure}

In Table \ref{tab:ansurf} we present
the results of TBGGA calculations on $(001)$ and $(110)$ slabs of
two thicknesses. We have first checked the convergence of the
magnetic anisotropy with respect to the number of $k$-points in
the first Brillouin zone. It is seen in Fig. \ref{fig:conv_kpoints} that a good
convergence is obtained above 1000 $k$-points.
It is found that for both orientations the easy
axis is perpendicular to the surface plane. The two surfaces, however, have a
different behavior with respect to in-plane magnetization. For the
$(001)$ surface the energy is changing by only $4.10^{-5}$eV
between a magnetic configuration along  the square edge
($\varphi=0$) and along the diagonal of the square
($\varphi=\pi/4$). This confirms our previous symmetry analysis
which predicts  no in-plane dependence of the energy  at second
order in perturbation. In the case of the $(110)$ slab with 11
layers the in-plane energy variation is almost one order of
magnitude larger and plays in favor of the second nearest neighbor
atomic direction ($\varphi=\pi/2$). Consequently the anisotropy is
very small in the $yz$ plane. Let us also point out that isolated
monolayers show a more pronounced out of plane anisotropy while
the in-plane energy profile seems to be less corrugated than for
thicker layers in the case of the $(110)$ orientation.

\begin{table}
\begin{tabular}{|c|c|c|c|c|}
\hline
Surface & $n_l$ & $n_k$ & $E_{tot}(\pi/2,0)-E_{tot}(0,0)$ & $E_{tot}(\pi/2,\pi/4)-E_{tot}(\pi/2,0)$ \\
\hline
$(001)$ &  1  & 40000 &  1.29meV          &   0.049meV  \\
\hline
$(001)$ &  11  & 1024 &  0.45meV          &   0.041meV  \\
\hline  \hline
      &        &      & $E_{tot}(\pi/2,0)-E_{tot}(0,0)$ & $E_{tot}(\pi/2,\pi/2)-E_{tot}(\pi/2,0)$ \\
\hline
$(110)$ &  1  & 40000 &  0.86meV          &   -0.087meV  \\
\hline
$(110)$ &  11  & 1024 &  0.19meV          &   -0.17meV  \\
\hline
\end{tabular}
\caption{Magneto-crystalline anisotropy, $E_{tot}(\theta,\varphi)$
being the total energy (per surface unit cell) corresponding to a
magnetization direction defined by the angles $\theta,\varphi$
with respect to the crystal axes (see text) for slabs of bcc Fe
with $(001)$ and $(110)$ orientations: $n_l$ is the number of layers
and $n_k$ is the number of $k$ points in the first Brillouin zone
used in the calculation.}
\label{tab:ansurf}
\end{table}

\subsection{Orbital moment}

In presence of a surface the coordination is reduced and the
symmetry is lowered leading to an enhancement of the spin moments
as discussed above. At surfaces the orbital moment is  also
enhanced as demonstrated in several theoretical and
experimental \cite{Eriksson91,Tischer95}
works. Our calculations on $(001)$ and $(110)$ slabs show that the
component $<L_{iz''}>$ of the orbital moment on the magnetization
direction is noticeably increased when $i$ belongs to the
outermost layer, especially for the (001) surface. On the
second layer this component has almost recovered its bulk value
for the $(110)$ surface whereas  oscillations occur for the
$(001)$ surface similarly to the behavior of the spin moment.
Finally $<L_{iz''}>$, contrary to the spin moment, depends sensitively
on the direction of the magnetization with the same type of laws
(Eqs.\ref{eq:k001} and  \ref{eq:k110}) as the magneto-crystalline
anisotropy. However it is found that for $\theta=\pi/2$ the
$\varphi$ dependence predicted by Eq.\ref{eq:k110} for the $(110)$
surface is almost negligible. On the contrary the variation of
$<L_{iz''}>$ with $\theta$ is noticeable and, in this respect, the
two surfaces behave differently: when the magnetization direction is
rotated from $\theta=0$ to $\theta=\pi/2$,  $<L_{iz''}>$ decreases
for the $(001)$ surface while it increases for the $(110)$ one. In
addition, this variation is larger in absolute value on the $(001)$
than on the (110) surface.

Finally it is expected that equation \ref{eq:rap_ani}  should be more obeyed
for $(001)$ surface than for the $(110)$ since the spin moment at the
outermost plane is saturated for the former and not for the latter.
Assuming that the contribution to the magnetocrystalline energy
comes from the outermost plane only, the ratio of the magnetocrystalline
anisotropy to that of the orbital moment for surface atoms, has the
wrong sign for $(110)$, while for the $(001)$ surface the sign is correct
but the ratio is around three times smaller than $\xi/4$. Indeed the
exchange splitting of the $(001)$ surface is smaller than the $d$
bandwidth and the contribution of spin-flip excitations to the
magnetocrystalline anisotropy cannot be neglected.

\begin{figure}[!fht]
\begin{center}
\includegraphics*[scale=0.4,angle=0]{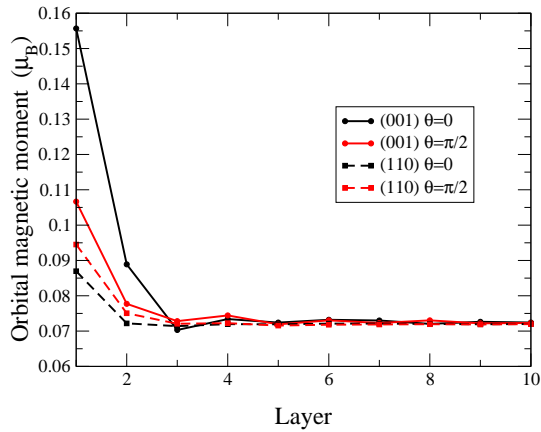}
\end{center}
\caption{Variation of the component of the orbital magnetic moment
on the magnetization direction (per atom) as a function of the
atomic layer in the $(110)$ and $(001)$ slabs (20 layers) for a
magnetization perpendicular ($\theta=0$) or parallel to the
surface ($\theta=\pi/2, \varphi=0$). }
\label{fig:surf_orbital_moment}
\end{figure}

\section{Study of the monatomic wire}

Although the study of the unsupported monatomic wire, i.e., a
periodic linear chain of identical atoms with a single atom per
unit cell, is somewhat academic it is however an interesting
object for the following reasons: i) it can be used as a model
since analytical TB results can be derived which are useful for a
theoretical understanding and a direct identification of the
orbital character of the bands obtained in ab-initio calculations,
ii) it also allows to investigate how a reduced dimensionality may modify
magnetism in ferromagnetic metals\cite{Mokroussov05} or induce it in non-magnetic
materials\cite{Delin04a,Delin04b,Delin04c}, iii) last, but not least, such objects,
several atom long, have been observed in break junction
experiments \cite{Smit01} (unfortunately not for Fe, Co or Ni).

\subsection{Non magnetic band structure of the monatomic wire}

\begin{figure}[!fht]
\begin{center}
\includegraphics*[scale=0.4,angle=0]{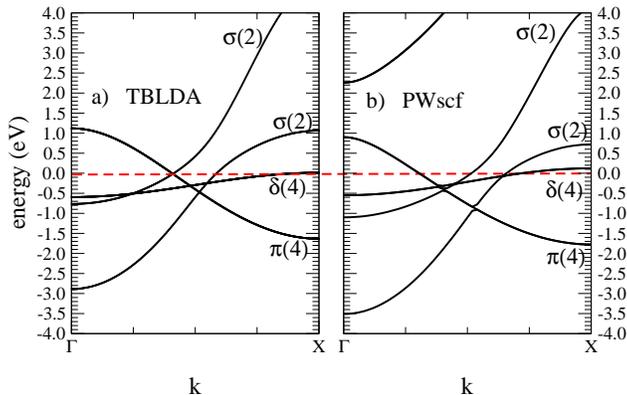}
\end{center}
\caption{a) TBLDA and b) PWscf band structure of a non-magnetic Fe
monatomic wire with an interatomic distance of 4.29 a.u.. Each
band is labelled by its symmetry character and degeneracy (including spin). }
\label{fig:wire_bands_nomag}
\end{figure}

The band structure of the non magnetic Fe monatomic wire,
neglecting spin-orbit coupling, obtained from the TBLDA model and
from the ab-initio PWscf code in the GGA are shown in Fig.\ref{fig:wire_bands_nomag} for an
interatomic distance of 4.29 a.u.(2.27\AA), {\sl i.e.}, at the
equilibrium distance predicted from the spin polarized PWscf GGA
calculations. It can be seen that except for the upper band in the
PWscf band structure, the agreement is satisfactory.

Let us now identify the character of the bands by using the
simplest TB model, i.e., in which overlap is neglected and hopping
integrals are restricted to first nearest neighbors. We take the
$z$ axis along the chain. For a given spin the ($9\times 9$)
hopping matrix can be rearranged into five square blocks on the
diagonal: the first one involves $s, p_z$ and $d_{3z^2-r^2}$
orbitals, the second and third ones are identical and involve
($p_x, d_{zx}$) and $(p_y, d_{yz})$ orbitals, respectively,
finally the fourth and fifth ones are also identical and
correspond to the $d_{xy}$ and $d_{x^2-y^2}$ orbitals,
respectively. Consequently, taking into account spin degeneracy,
the band structure consists in three two-fold degenerate bands of
symmetry $\sigma$, two four-fold degenerate bands of symmetry
$\pi$ and one four-fold degenerate band of symmetry $\delta$.

The $\delta$ band is easily identified as being the flattest band
which disperses positively. Indeed its dispersion relation is that
of a linear chain of $d_{xy}$ (or $d_{x^2-y^2}$) orbitals with a
small and negative hopping integral $dd\delta$. In Fe the $p$
level is much higher in energy than the $d$ level. As a
consequence, the mixing between $p$ and $d$ orbitals is small and
the $\pi$ bands split into a lower band with an almost pure
$d_{zx}$ (or $d_{yz})$ character and a higher one with an almost
pure $p_x$ or $p_y$ character. The first one shows a dispersion
very similar to that of a linear chain of $d_{zx}$ (or $d_{yz})$
orbitals, i.e., it disperses negatively since the corresponding
integral $dd\pi$ is positive. The second one disperses positively
since $pp\pi$ is negative, this band is present in the PWscf
calculation, but is outside the energy range of Fig.\ref{fig:wire_bands_nomag} in the TBLDA
results. This means that the TB parameters relative to $p$ bands
are not very accurate but since the higher $\pi$ band is
unoccupied, this inaccuracy will have no influence on our results
for the ground state. The two remaining bands are the two lowest
$\sigma$ bands. An analysis of the character of these bands using
the TBLDA model shows that the lowest band has almost no $p$
character, the weight of $s$ and $d_{3z^2-r^2}$ orbitals are
almost the same at the $\Gamma$ point while at the $X$ point, the
state is almost a pure $d$ one. For the next $\sigma$ band the
$d_{3z^2-r^2}$ character decreases continuously from $\simeq 0.5$
to $0$ along $\Gamma X$, the $p_z$ character increases
continuously from 0 to 1 while the $s$ character is $\simeq 0.5$
at $\Gamma$, has a maximum at the midpoint and vanishes at the $X$
point.

\subsection{Magnetic transition}

\begin{figure}[!fht]
\begin{center}
\includegraphics*[scale=0.4,angle=0]{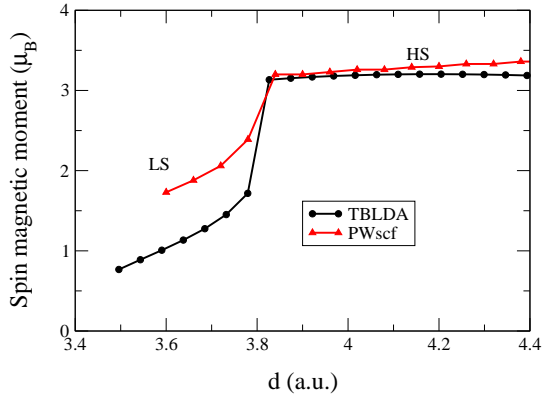}
\end{center}
\caption{TBLDA and PWscf spin moment of a monoatomic wire, as a function of the interatomic distance. }
\label{fig:M_d_wire}
\end{figure}

As in the bulk we have studied the appearance of a magnetic moment
when the interatomic distance $d$ increases using the TBLDA model
as well as the PWscf code with GGA. As seen in
Fig.\ref{fig:M_d_wire} a low spin state (LS) is found at short
interatomic distances with a slowly increasing moment, then, around
$d=3.8$ a.u., the moment increases abruptly to reach a high spin
state (HS) corresponding to the saturated configuration. The
agreement between both methods is quite satisfactory, in
particular, the HS state appears at the same interatomic distance.
It should however be mentioned that the TBGGA fails to reproduce
the wire band structure in the range of interatomic distances
involved at the transition due to an incorrect position of the $s$
level. Actually, for interatomic distances around 3.8 a.u. this
$s$ level is pushed at higher energies and instead of a LS/HS
transition one finds a NM/HS transition. Let us recall that the TB
parameters are fitted on bulk results only, and for interatomic
distances and coordinence larger than in the wire. Therefore, the
extrapolation of the law giving the atomic levels as a function of
the atomic environment may fail for the wire. From the above study
it appears that the TBLDA levels are much better than the GGA
ones. As a consequence, in the following, LDA parameters will be
used.

\subsection{Spin-orbit coupling effects}

\subsubsection{Effects of spin-orbit coupling on the band structure from perturbation theory.}

The removal of degeneracies due to the spin-orbit coupling can be
predicted by perturbation theory. As seen in Fig.\ref{fig:wire_bands_nomag} two types of
degeneracies occur for the wire: in addition to spin degeneracy,
either a band is degenerate for
any value of {\bf k} due to symmetry or the degeneracy is limited
to band crossings. In the former case $H_{\text{so}}$ may produce a
splitting of the bands while in the latter it may open a gap
between the two crossing bands.

Let us first consider the non magnetic case for which the band
structure should be independent of $\theta$ and $\varphi$. The
$\delta$ band is four-fold degenerate. For $\theta=0 (\pi/2)$
$H_{\text{so}}$ couples the $d_{xy}$ and $d_{x^2-y^2}$ orbitals with the
same (opposite) spins with a coupling matrix element $\pm i\xi$ (see Appendix A).
Consequently the $\delta$ band is unfolded symmetrically with a
band splitting given by $2\xi$. The same type of arguments applies
to the lowest four-fold $\pi$ band if the small $p$ character is
neglected but the coupling matrix element is now $\pm i\xi/2$ and
therefore the band splitting is equal to $\xi$. No removal of
degeneracy is expected on the two-fold $\sigma$ bands since there
is no matrix elements of $H_{\text{so}}$ between the $d_{3z^2-r^2}$
orbitals of opposite spins. Finally in the unperturbed band
structure (Fig.\ref{fig:wire_bands_nomag}) there are two crossing points with the $\sigma$
and $\delta$ bands which are not coupled by $H_{\text{so}}$. Thus there
is no removal of degeneracy. On the contrary the other crossing
points (between $\sigma$ and $\pi$ bands, or $\pi$ and $\delta$
bands) are avoided.

In the magnetic case the band structure is independent of
$\varphi$ for symmetry reasons but depends on the polar angle
$\theta$ when spin-orbit coupling is taken into account. This is
the consequence of the removal of spin degeneracy in the
unperturbed state. Indeed for $\theta=0$, $H_{\text{so}}$ couples states
with the same spin and a splitting $2\xi (\xi)$ exists in the
$\delta$ (lowest $\pi$) bands, for both spins, while for
$\theta=\pi/2$, the bands of up and down spins being well
separated in energy, $H_{\text{so}}$ has a negligible effect on the
$\delta$ and lowest $\pi$ bands. Finally, removals of degeneracy
are expected at some crossing points. The number of opened gaps
should be small at $\theta=0$ since, for bands of different
symmetries, only states with opposite spins may be coupled by
$H_{\text{so}}$. As the exchange splitting is large the number of avoided
crossings should be very small. On the contrary at $\theta=\pi/2$
a detailed analysis of the $H_{\text{so}}$ matrix (see Appendix A)
reveals that states with the same as well as different spins may
be coupled and the number of avoided crossings is expected to be
larger than for $\theta=0$.

\begin{figure}[!fht]
\begin{center}
\includegraphics*[scale=0.4,angle=0]{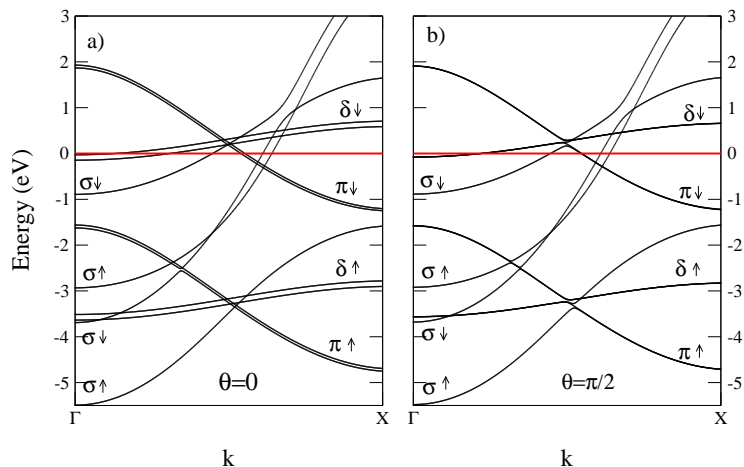}
\end{center}
\caption{TBLDA band structure including spin-orbit coupling for a
magnetic Fe monatomic wire (interatomic distance d=4.16 a.u.) with
a magnetization a) parallel and b) perpendicular to the wire.}
\label{fig:wire_bands_S0}
\end{figure}

The TBLDA calculation is presented in Fig.
\ref{fig:wire_bands_S0} for a magnetization parallel ($\theta=0$)
or perpendicular ($\theta=\pi/2$) to the wire. The results are in
perfect agreement with the predictions of perturbation theory.
Note that a rather good overall agreement is also obtained with
PWscf calculations \cite{Viret06}. In particular the splitting of
the $\delta$ band in the latter calculations is 120meV, i.e.,
exactly $2\xi$ which confirms that spin-orbit coupling is an intra-atomic
effect and that $\xi$ is a purely atomic
quantity, i.e., it does not depend on the atomic environment.


\subsubsection{Magnetocrystalline anisotropy.}

Since the monatomic wire is a one dimensional system with the
lowest possible coordination we expect a magnetocrystalline
anisotropy larger than at the surface. Moreover, due to the axial
symmetry of the wire the energy will only depend on the angle
$\theta$ between the magnetization and the axis of the wire. By
imposing the symmetry properties of the wire to the quadratic form
in $l, m,$ and $n$ giving the magnetocrystalline anisotropy in
second order perturbation theory, we find:

\begin{equation}
\Delta E^{(2)}_i(\theta,\varphi)-\Delta E^{(2)}_i(0,0)=K'_1\sin^2\theta
\end{equation}

\begin{figure}[!fht]
\begin{center}
\includegraphics*[scale=0.4,angle=0]{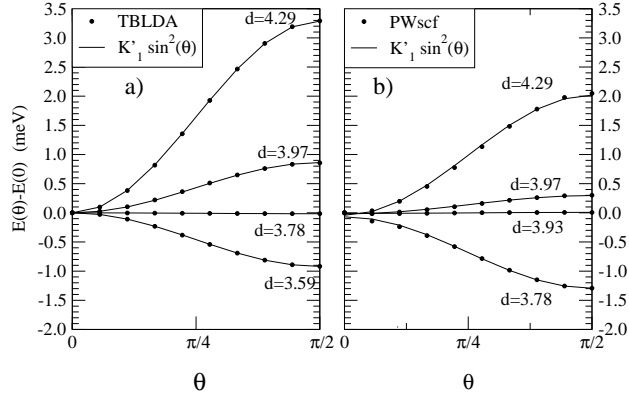}
\end{center}
\caption{Variation of the magnetocrystalline anisotropy (per
atom) of a monatomic wire as a function of the angle $\theta$ from
a TBLDA calculation (left panel)  and PWscf calculation (right
panel) for various interatomic distances $d$ (in a.u.).}
\label{fig:wire_magnetic_anisotropy}
\end{figure}

In Fig. \ref{fig:wire_magnetic_anisotropy} we present the results
of our TBLDA and PWscf calculations obtained at various
interatomic distances. The variation of energy with $\theta$ is
perfectly fitted by the above equation. Interestingly an inversion
of the easy axis is observed (i.e., a change of sign of
$K'_1$). Indeed, for interatomic distances $d$ larger than 3.78
a.u. (3.93 a.u. with PWscf) the easy axis is along the wire, while
it is perpendicular when $d$ is smaller. This inversion of the
easy axis occurs at the LS/HS magnetic transition. A detailed
analysis of the charge distribution shows that the magnetic
transition is accompanied by a noticeable change of filling of the
$\delta$ bands.

\subsubsection{Orbital moment.}

Let us first derive the expressions of the components of the
orbital moment relative to the spin framework using perturbation
theory. In a monatomic wire of atoms with only $d$ orbitals, the
band hamiltonian is reduced to five decoupled linear chains with a
single orbital on each atom, thus
 $\mathcal{N}_{\lambda\lambda'\sigma}({\bf k},E)=
 \mathcal{N}_{\lambda\lambda\sigma}({\bf
 k},E)\delta_{\lambda\lambda'}$ and Eq.\ref{eq:LNN} becomes:

\begin{equation}
<{\bf
L}^{''}_i>=-2\xi\sum_{\lambda\mu}\sum_{\sigma}
\bra{\bar{\lambda}\sigma}{\bf L}^{''}_i\ket{\bar{\mu}\sigma}
\bra{\bar{\lambda}\sigma}{\bf L}.{\bf S}\ket{\bar{\mu}\sigma}^*I_{\lambda\mu}
\end{equation}

\noindent where $I_{\lambda\mu}$ denotes the term in
Eq.\ref{eq:LNN} involving the summation over ${\bf k}$ but with
$\lambda'=\lambda$ and $\mu'=\mu$.

In order to find the angular dependence of $<{\bf L}^{''}_i>$, we
use Eq.\ref{eq:L} with the matrix elements of ${\bf L}_i$ and of
$H_{so}$ given in Appendix A taking $\varphi=0$, for simplicity,
since the wire has an axial symmetry. Then it is easily seen that:

\begin{eqnarray}
<L_{ix^{''}}>&=&K_{1x^{''}}\sin2\theta  \\
<L_{iz^{''}}>&=&K_{0z^{''}}+K_{1z^{''}}\sin^2\theta
\end{eqnarray}

\noindent while $<L_{iy^{''}}>=0$ as expected since $y^{''}$ is
perpendicular to the plane defined by the wire and the spin
quantization axis.

\begin{figure}[!fht]
\begin{center}
\includegraphics*[scale=0.4,angle=0]{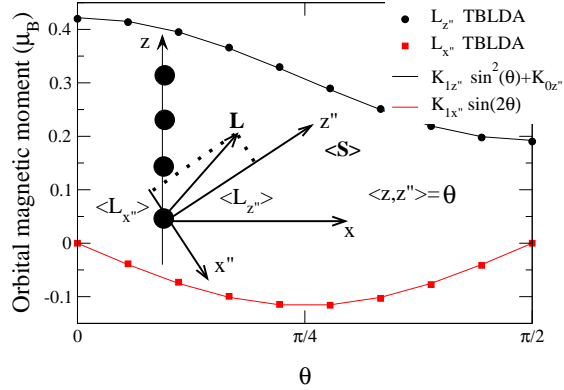}
\end{center}
\caption{Variation of the components of the orbital moment
relative to the spin framework for a monatomic wire (interatomic
distance $d=4.29$ a.u.) as a function of the direction of
magnetization given by the angle $\theta$ obtained with TBLDA
parameters.} \label{fig:wire orbital moment}
\end{figure}

The results of the TBLDA calculations of $<L_{ix^{''}}>$ and
$<L_{iz^{''}}>$, shown in Fig.\ref{fig:wire orbital moment} for an
interatomic distance $d=4.29$ a.u., are in perfect agreement with
this analysis. Note that $<L_{iz^{''}}>$ can reach a value as
large as $0.42\mu_B$ when the magnetization is along the wire and
that the anisotropy of the orbital moment
$<L_{iz^{''}}(\theta=\pi/2)>-<L_{iz^{''}}(\theta=0)>=-0.22\mu_B$,
is rather large. At the same interatomic distance (see
Fig.\ref{fig:wire_magnetic_anisotropy}) the corresponding
magnetocrystalline anisotropy is 3.3.meV. Thus the ratio of this
anisotropy to that of the orbital moment is equal to -15meV which
is in perfect agreement with Eq.\ref{eq:rap_ani} ($-\xi/4=-15meV$). This was
rather expected since the spin up $d$ bands are completely filled
and well separated ($\simeq 3eV$) from the spin down ones at this
interatomic distance, both bands being much narrower than at the surface.

At shorter distances for which the LS state is found,
Eq.\ref{eq:rap_ani}  is no longer valid and even does not predict
the sign of this ratio since the magnetocrystalline anisotropy
changes sign, contrary to that of the orbital moment.

 Finally let us note that in the presence
of spin-orbit coupling the spin moment is very slightly anisotropic:
typically $<S_{z"}>$ varies by about 0.02$\mu_B$ between $\theta=0$ and
$\theta=\pi/2$. Moreover, similarly to the orbital moment, a small
component $<S_{x"}>$ is present when $\theta\in]0,\pi/2[$.

\section{Conclusion}

We have shown that starting from the parametrized $spd$ TB model set up
by Mehl and Papaconstantopoulos \cite{Mehl96},  adding a Stoner-like spin polarization
term and a spin-orbit coupling term, i.e, introducing only two additional
parameters $I_{dd'}$ and $\xi$, we have been able to describe in detail the
magnetic properties (spin moments, orbital moments and magnetocrystalline anisotropy
energy) of iron in systems of various dimensionalities and coordinences. Whenever possible
the results have been compared with those of the PWscf code or other existing theoretical
or experimental data, and the agreement is excellent.
Our simple TB model allowed us to derive some new results. For example in the case
of surfaces we have studied the variation of the component of the orbital magnetic
moment on the spin quantization axis $<L_{z"}>$ as a function of depth, and shown that
the enhancement of this quantity is still noticeable on the second layer for the $(001)$ surface but
almost cancels on the third and innermost layers. In the wire we have found that
the easy axis of magnetization is along the wire at the theoretical equilibrium distance
but can switch to the perpendicular direction under compression. Moreover 
$<L_{z"}>$ is strongly enhanced and highly anisotropic since for $z"$ along the wire
it is about $0.42\mu_B$ and decreases to half this value for the perpendicular direction.
In addition for intermediate orientations the orbital moment has a non-negligible
component $<L_{x"}>$ perpendicular to $z"$ and in the plane made by the wire and $z"$. This
component is also anisotropic and proportional to $\sin 2\theta$ as predicted by perturbation
theory. Finally we have shown that the law of Bruno\cite{Bruno89} relating the MAE and the anisotropy 
of the orbital moment, is perfectly obeyed for the wire at equilibrium distance,
which is not the case for the surfaces, even for the $(001)$ orientation at which the
magnetic moment is saturated.

The success of our model opens up the possibility of obtaining accurate results on other
elements and systems with much more complex geometries.

\acknowledgments

\noindent
It is our pleasure to thank M. Viret for very fruitful discussions and 
A. Dal Corso and A. Smogunov for precious advice concerning the PWscf code.

\appendix

\section{The matrix elements of the orbital moment and spin-orbit coupling operators}

The matrix elements $\bra{\bar{\lambda}\sigma}{\bf L}.{\bf S}\ket{\bar{\mu}\sigma'}$,
where $\bar{\lambda}$ and $\bar{\mu}$ are
the angular parts of $d$ orbitals $(d_{xy}, d_{yz}, d_{zx}, d_{x^2-y^2},
d_{3z^2-r^2})$ centered at the same site as the
angular momentum operator, can be easily calculated as a function
of the direction (defined by the polar and azimuthal angles
$\theta$ and $\varphi$ relative to the crystal axes) of the spin
quantization axis $z''$ which is often taken in the magnetization
direction that may be different from the $z$ axis of the crystal.
This is achieved by choosing a new coordinate framework $x''$,
$y''$, $z''$ referred to as the spin framework which is obtained
by rotating first the $xy$ axes of the crystal by the angle
$\varphi$ around $z$. This gives a new framework $x', y', z'$
which is then rotated by the angle $\theta$ around $y'$. Using
the matrix elements of {\bf L} in the basis of real d orbitals:

\begin{displaymath}
\bra{\lambda\sigma}L_x\ket{\mu\sigma}= \left(\begin{array}{ccccc}
0&0&-i&0&0 \\
0&0&0&-i&-i\sqrt{3} \\
i&0&0&0&0 \\
0&i&0&0&0  \\
0&i\sqrt{3}&0&0&0
\end{array}\right)
\end{displaymath}

\begin{displaymath}
\bra{\lambda\sigma}L_y\ket{\mu\sigma}= \left(\begin{array}{ccccc}
0&i&0&0&0 \\
-i&0&0&0&0 \\
0&0&0&-i&i\sqrt{3} \\
0&0&i&0&0  \\
0&0&-i\sqrt{3}&0&0
\end{array}\right)
\end{displaymath}

\begin{displaymath}
\bra{\lambda\sigma}L_z\ket{\mu\sigma}= \left(\begin{array}{ccccc}
0&0&0&2i&0 \\
0&0&i&0&0 \\
0&-i&0&0&0 \\
-2i&0&0&0&0  \\
0&0&0&0&0
\end{array}\right)
\end{displaymath}

\noindent one finds:

\begin{displaymath}
\bra{\bar{\lambda}\uparrow}{\bf L}.{\bf S}\ket{\bar{\mu}\uparrow}=
\left(\begin{array}{ccccc} 0&\frac{i}{2}\sin \varphi\sin \theta&
\frac{-i}{2}\cos \varphi\sin \theta&i\cos \theta&0 \\
\frac{-i}{2}\sin \varphi\sin \theta&0&\frac{i}{2}\cos \theta&
\frac{-i}{2}\cos \varphi\sin \theta&\frac{-i\sqrt{3}}{2}\cos
\varphi\sin \theta \\
\frac{i}{2}\cos \varphi\sin \theta&\frac{-i}{2}\cos \theta&0&
\frac{-i}{2}\sin \varphi\sin \theta&\frac{i\sqrt{3}}{2}\sin
\varphi\sin \theta \\
-i\cos \theta&\frac{i}{2}\cos \varphi\sin \theta&\frac{i}{2}\sin
\varphi\sin \theta&0&0  \\
0&\frac{i\sqrt{3}}{2}\cos \varphi\sin
\theta&\frac{-i\sqrt{3}}{2}\sin \varphi\sin \theta&0&0
\end{array}\right)
\end{displaymath}

\begin{displaymath}
\bra{\bar{\lambda}\uparrow}{\bf L}.{\bf S}\ket{\bar{\mu}\downarrow}=
\left(\begin{array}{ccccc}
0&\frac{1}{2}f(\varphi,\theta)&\frac{1}{2}g(\varphi,\theta)&-i\sin\theta&0 \\
-\frac{1}{2}f(\varphi,\theta)&0&\frac{-i}{2}\sin\theta&\frac{1}{2}g(\varphi,\theta)&
\frac{\sqrt{3}}{2}g(\varphi,\theta)  \\
-\frac{1}{2}g(\varphi,\theta)&\frac{i}{2}\sin\theta&0&-\frac{1}{2}f(\varphi,\theta)
&\frac{\sqrt{3}}{2}f(\varphi,\theta)  \\
i\sin\theta&-\frac{1}{2}g(\varphi,\theta)&\frac{1}{2}f(\varphi,\theta)&0&0  \\
0&-\frac{\sqrt{3}}{2}g(\varphi,\theta)&-\frac{\sqrt{3}}{2}f(\varphi,\theta)&0&0
\end{array}\right)
\end{displaymath}

\noindent where $f(\varphi,\theta)=\cos \varphi+i\sin
\varphi\cos\theta$ and $g(\varphi,\theta)=\sin \varphi-i\cos
\varphi\cos\theta$. The other blocks of the $(10\times 10)$
spin-orbit matrix are obtained from the relations:

\begin{eqnarray}
\bra{\bar{\lambda}\downarrow}{\bf L}.{\bf S}\ket{\bar{\mu}\uparrow}&=&-
\bra{\bar{\lambda}\uparrow}  {\bf L}.{\bf S}\ket{\bar{\mu}\downarrow}^* \\
\bra{\bar{\lambda}\downarrow}{\bf L}.{\bf S}\ket{\bar{\mu}\downarrow}&=&
\bra{\bar{\lambda}\uparrow}  {\bf L}.{\bf S}\ket{\bar{\mu}\uparrow}^*
\end{eqnarray}

\noindent in which $*$ denotes the complex conjugate. In addition a very useful
relation has been derived by Bruno\cite{Bruno89th}:

\begin{equation}
\text{Re}[
\bra{\bar{\lambda}\uparrow}{\bf L}.{\bf S}\ket{\bar{\mu}\downarrow}
\bra{\bar{\mu}'\downarrow}{\bf L}.{\bf S}\ket{\bar{\lambda}'\uparrow}
 ]+
\bra{\bar{\lambda}\uparrow}{\bf L}.{\bf S}\ket{\bar{\mu}\uparrow}
\bra{\bar{\mu}'\uparrow}{\bf L}.{\bf S}\ket{\bar{\lambda}'\uparrow}
 =\text{Cst.}
\end{equation}

Let us note however that the spin quantization axis could have
been taken along the $z$ axis of the crystal, in which case the
spin-orbit matrix elements would be given by the above matrices
with $\theta=\varphi=0$ but the spin polarized term of 
the Hamiltonian becomes:

\begin{displaymath}
-\frac{1}{4}\big[(\Delta_{\lambda}^i +\Delta_{\mu}^j)\mathcal{S}_{ij}^{\lambda \mu}\big]\otimes \left(
\begin{array}{cc}
\cos\theta & \exp(-i\varphi)\sin\theta \\
\exp(i\varphi)\sin\theta & -\cos\theta
\end{array} \right)
\end{displaymath}

\noindent where $\otimes$ means the direct product of matrices.

In the case of collinear spins and in the presence of spin-orbit
coupling the first point of view is more convenient to treat
spin-orbit coupling effects within perturbation theory since
$H_{\text{so}}$ describes the perturbation completely. The second
point of view is preferable when dealing with non-collinear spins.
Indeed this avoids the transformation of the inter-atomic part of
$H_{TBHF}$ since, in that case, the spin functions are not the same
at the two sites.

\section{Calculation of the orbital moment for a non-orthogonal
basis set}

Let us generalize Eq.\ref{eq:Lseci} to take overlap into account.
If we note that the integral in Eq.\ref{eq:Lric} gives the
population of the spin-orbital $\ket{ilm\sigma}$, an obvious
generalization is to replace this population by the Mulliken one.
Thus $\rho_{ilm\sigma}(E)$ becomes:

\begin{equation}
\rho_{ilm\sigma}(E)=\text{Re}\sum_{\substack{i'l'm' \\ n}}a^{n*}_{ilm\sigma}a^n_{i'l'm'\sigma}
\mathcal{S}_{ii'}^{lm,l'm'}\delta(E-E_n)
\end{equation}

\noindent with:

\begin{equation}
\ket{\psi_n}=\sum_{ilm\sigma}a^n_{ilm\sigma}|ilm\sigma>
\end{equation}

\noindent and

\begin{equation}
\mathcal{S}_{ii'}^{lm,l'm'}=\braket{ilm\sigma}{i'l'm'\sigma}
\end{equation}

\noindent thus

\begin{equation}
<L_{iz''}>=\text{Re}
\sum_{\substack{lm,i'l'm'\sigma\\ n \text{ occ}}}ma^{n*}_{ilm\sigma}a^n_{i'l'm'\sigma} \mathcal{S}_{ii'}^{lm,l'm'}
\end{equation}

\noindent and, after simple algebraic manipulations:

\begin{equation}
<L_{iz''}>=\text{Re}
\sum_{\substack{lm,i'l'm'\sigma \\ n \text{ occ}}}
\braket{\psi_n}{ilm\sigma}m[\mathcal{S}^{-1}]_{ii'}^{lm,l'm'}\braket{i'l'm'\sigma}{\psi_n}
\end{equation}

The generalization of Eq.\ref{eq:Lvec} yields:

\begin{equation}
<{\bf L}^{''}_i>=\text{Re}
\sum_{\substack{lm,l''m'',i'l'm'\sigma \\ n\text{ occ}} }
\braket{\psi_n}{ilm\sigma}
[{\bf L}^{''}_i]_{lm,l''m''}[\mathcal{S}^{-1}]_{ii'}^{l''m'',l'm'}
\braket{i'l'm'\sigma}{\psi_n}
\end{equation}

\noindent and, in the basis $\ket{i\lambda\sigma}$:

\begin{equation}
<{\bf L}^{''}_i>=\text{Re}
\sum_{\substack{\lambda,\mu,i'\nu\sigma \\ n \text{ occ}}}
\braket{\psi_n}{i\lambda\sigma}
[{\bf L}^{''}_i]_{\lambda\mu}[\mathcal{S}^{-1}]_{ii'}^{\mu\nu}
\braket{i'\nu\sigma}{\psi_n}
\end{equation}

\noindent with:

\begin{equation}
\braket{i'\nu\sigma}{\psi_n}=\sum_{i\lambda\sigma}c^n_{i\lambda\sigma}\mathcal{S}^{\nu\lambda}_{i'i}
\end{equation}

\noindent so that we get finally:

\begin{equation}
<{\bf L}^{''}_i>=\text{Re}
\sum_{\substack{\lambda\mu i'\nu\sigma \\ n \text{ occ}}}
c^{n*}_{i\lambda\sigma}[{\bf L}^{''}_i]_{\lambda\mu}\mathcal{S}^{\mu\nu}_{ii'}c^n_{i'\nu\sigma}
\end{equation}

\end{document}